\documentclass[11pt,a4paper,onecolumn]{IEEEtran}

\usepackage[ breaklinks,
             colorlinks = true,
             linkcolor = blue,
             urlcolor  = blue,
             citecolor = red,
             anchorcolor = green,
]{hyperref}

\usepackage{amssymb,amsmath,amsfonts,amsthm}
\usepackage{color}
\usepackage[T1]{fontenc}

\theoremstyle{remark}
\newtheorem{axiom}{Axiom}
\theoremstyle{plain}
\newtheorem{lemma}{Lemma}
\newtheorem{corollary}[lemma]{Corollary}
\newtheorem{theorem}[lemma]{Theorem}
\newtheorem{prop}[lemma]{Proposition}
\newtheorem{definition}[lemma]{Definition}

\newcommand{\cX}{\mathcal{X}}
\newcommand{\cY}{\mathcal{Y}}
\newcommand{\cZ}{\mathcal{Z}}
\newcommand{\cP}{\mathcal{P}}
\newcommand{\cT}{\mathcal{T}}
\newcommand{\cQ}{\mathcal{Q}}

\renewcommand{\d}{\mathrm{d}}
\newcommand{\sym}{\textnormal{sym}}
\newcommand{\Rth}{R_{\rm th}}

\newcommand{\ua}{\textnormal{\tiny$\uparrow$}}
\newcommand{\uua}{\textnormal{\tiny$\uparrow\!\uparrow$}}
\newcommand{\uuua}{\textnormal{\tiny$\uparrow\!\uparrow\!\uparrow$}}
\newcommand{\uuda}{\textnormal{\tiny$\uparrow\!\uparrow\!\downarrow$}}
\newcommand{\ddda}{\textnormal{\tiny$\downarrow\!\downarrow\!\downarrow$}}
\newcommand{\da}{\textnormal{\tiny$\downarrow$}}
\newcommand{\dda}{\textnormal{\tiny$\downarrow\!\downarrow$}}
\newcommand{\uda}{\textnormal{\tiny$\uparrow\!\downarrow$}}

\DeclareMathOperator*{\argmin}{\arg\min}
\DeclareMathOperator*{\relint}{relint}

\newcommand{\poly}{\mathrm{poly}}

\newcommand{\hypo}[3]{
\begin{align}
  \textnormal{#1} : \quad \qquad \textnormal{{null hypothesis}:} \quad & \textnormal{#2} \,, \nonumber \\
  \textnormal{{alternative hypothesis}:} \quad & \textnormal{#3} .  
\end{align}
}
\newcommand{\hypoinformal}[2]{
\begin{align}
  \textnormal{{null hypothesis}:} \quad & \textnormal{#1} \,, \nonumber \\
  \textnormal{{alternative hypothesis}:} \quad & \textnormal{#2} .  
\end{align}
}
\newcommand{\hypoinformalc}[2]{
\begin{align}
  \textnormal{{null hypothesis}:} \quad & \textnormal{#1} \,, \nonumber \\
  \textnormal{{alternative hypothesis}:} \quad & \textnormal{#2} , 
\end{align}
}

\begin{document}
 
\title{Operational Interpretation of R\'enyi Information Measures via Composite Hypothesis Testing Against Product and Markov Distributions}
 
 \author{Marco Tomamichel$^{\dag}$, {\em Senior Member, IEEE}   \hspace{.025in}  and \hspace{.025in}  Masahito Hayashi$^{\ddag*}$, {\em Fellow, IEEE} 
 \thanks{$^{\dag}$ Centre for Quantum Software and Information, University of Technology Sydney, Australia (Email: \url{marco.tomamichel@uts.edu.au})}  
 \thanks{$^\ddag$  Graduate School of Mathematics, Nagoya University, Japan (Email:
  \url{masahito@math.nagoya-u.ac.jp})}    
  \thanks{$^*$ Centre for Quantum Technologies, National University of Singapore (NUS), Singapore}  
   
  \thanks{This paper was presented in part at the 2015 International Symposium on Information Theory in Hong Kong, China. } }
   

\maketitle

\begin{abstract}
  We revisit the problem of asymmetric binary hypothesis testing against a composite alternative hypothesis. We introduce a general framework to treat such problems when the alternative hypothesis adheres to certain axioms. In this case we find the threshold rate, the optimal error and strong converse exponents (at large deviations from the threshold) and the second order asymptotics (at small deviations from the threshold). We apply our results to find operational interpretations of various R\'enyi information measures. In case the alternative hypothesis is comprised of bipartite product distributions, we find that the optimal error and strong converse exponents are determined by variations of R\'enyi mutual information. In case the alternative hypothesis consists of tripartite distributions satisfying the Markov property, we find that the optimal exponents are determined by variations of R\'enyi conditional mutual information.
  In either case the relevant notion of R\'enyi mutual information depends on the precise choice of the alternative hypothesis.
  As such, our work also strengthens the view that different definitions of R\'enyi mutual information, conditional entropy and conditional mutual information are adequate depending on the context in which the measures are used. 
\end{abstract}




\section{Introduction}


Let us first consider simple hypothesis testing.  
Here the null hypothesis states that a random variable $X^n$ follows the independent and identical (i.i.d.) law $P^{\times n}$ and 
the alternative hypothesis states that $X^n$ follows the i.i.d.\ law $Q^{\times n}$, where $P$ and $Q$ are probability mass functions on a finite set~$\cX$. We write this as follows:
\hypoinformal{$X^n \sim P^{\times n}$}{$X^n \sim Q^{\times n}$}

Assume now that our test $T^n$ is given as a randomized function from the observed event in $\cX^n$
to $\{0,1\}$. Here the values $1$ and $0$ signify that we accept and reject the null hypothesis, respectively.
We are particularly interested in the asymmetric case where two kind of errors are treated differently. The \emph{type-I error}, given as $\alpha_n = P^{\times n}[T^n(X^n) = 0]$, is the probability that the test rejects the null hypothesis when it is true. The \emph{type-II error}, given as $\beta_n = Q^{\times n}[T^n(X^n) = 1]$, is the probability that the test confirms the null hypothesis when the alternative hypothesis is true. (See Section~\ref{sec:binaryhypo} for formal definitions of these quantities.)

On the one hand, if we impose a constant constraint on the type-I error,
namely if we require that $\alpha_n \leq \varepsilon$ for some $\varepsilon \in (0,1)$, then
the there exists a sequence of tests such that $\beta_n$ goes zero exponentially fast in $n$.
The exponent is known to be the relative entropy, $D(P\|Q)$. This is Stein's lemma~\cite{chernoff56} (see also~\cite{dembo98,bucklew90}) and we also call this exponent the \emph{threshold rate} of the problem. (See Section~\ref{sec:opquant} for definitions of the relevant operational quantities.)
Further, Yushkevich~\cite{yushkevich53} derived the second order expansion (see also Strassen~\cite{strassen62} for a higher order expansion) of the optimal exponent as
\begin{align}
- \log \beta_n = n D(P\|Q) - \sqrt{n V(P\|Q)} \Phi^{-1}(\varepsilon) + o(\sqrt{n})\,, \label{eq:intro1}
\end{align}
where $\Phi$ is the cumulative standard normal distribution function and $V$ is the variance of the logarithmic likelihood ratio. (See Section~\ref{sec:renyi} for definitions of the relevant information quantities.) All our statements are independent of the basis choice for the logarithm as long as $\exp$ is taken as the inverse of $\log$, unless it is otherwise noted.

On the other hand, if we impose an exponential constraint on the type-II error, namely if we require that $\beta_n \leq \exp(-n R)$ for some rate $R \in (0, D(P\|Q))$, we find that the optimal type-I error decreases exponentially fast to zero with
\begin{align}
- \log \alpha_n = n 
\sup_{0 < s < 1} \frac{1-s}{s} \big(D_{s}(P\|Q)-R\big) + o(n) \,,   \label{eq:intro2}
\end{align}
where $D_{s}(P\|Q)$ is the R\'enyi relative entropy.
This is known as Hoeffding's bound~\cite{hoeffding65}, and the exponent is called \emph{error exponent} in the following.
Moreover, if the rate $R$ exceeds the threshold rate $D(P\|Q)$,
the minimum probability of the second error goes to $1$ exponentially fast
with
\begin{align}
- \log (1-\alpha_n) = n 
\sup_{s > 1} \frac{s-1}{s} \big(R - D_{s}(P\|Q)  \big) + o(n) \,.  \label{eq:intro3}
\end{align}
This is called \emph{strong converse exponent} in the following~\cite{csiszar71,han89}.\footnote{The form of~\eqref{eq:intro3} is due to Ogawa-Nagaoka~\cite{ogawa00}. Moreover, Nakagawa-Kanaya~\cite{nakagawa93} first treat the case of large $R$.}

These results can partially be extended to the case when the null hypothesis is composite (i.e.\ when the null hypothesis is comprised of a set of distributions) as a consequence of Sanov's theorem~\cite{sanov61}. In contrast, our goal is to extend the above results, in particular Eq.~\eqref{eq:intro1}--\eqref{eq:intro3}, to the setting where the alternative hypothesis is composite. More precisely, we want to consider a set $\cQ$ of distributions on $\cX$ and the maximal type-II error $\beta_n = \max_{Q \in \cQ} Q^{\times n}[T^n(X^n) = 1]$. We write the corresponding hypothesis testing problem as follows:
\hypoinformal{$X^n \sim P^{\times n}$}{$X^n \sim Q^{\times n}$ for some $Q \in \cQ$}

Sanov's theorem allows the alternative hypothesis to be
the set $\cQ = \{Q :\, D(P\|Q) > R\}$ for a given real number $R>0$.
More precisely,
when the first kind of error probability is restricted to $\alpha_n \leq \varepsilon$, 
the optimal exponent for $\beta_n$ is given as $- \log \beta_n = n R + o(n)$.
Moreover, the Hoeffding bound~\eqref{eq:intro2} was extended to 
certain classes of composite hypotheses which are composed of i.i.d.\ distributions~\cite{unnikrishnan11,shayevitz11}.

\subsubsection*{Our Contributions}
Our first main result establishes that, if the alternative hypothesis satisfies certain axioms discussed in Section~\ref{sec:axioms}, the above results, Eq.~\eqref{eq:intro1}-\eqref{eq:intro3}, hold as stated after we substitute
\begin{align}
  D(P\|Q) \to \min_{Q \in \cQ} D(P\|Q) = D(P\|\hat{Q}) , \quad V(P\|Q) \to V(P\|\hat{Q}), \quad D_s(P\|Q) \to \min_{Q \in \cQ} D_s(P\|Q),
\end{align}
where $\hat{Q} \in \cQ$ is the distribution that minimizes the relative entropy.
Hence, we generalize Stein's lemma, Yushkevich's second order expansion, the Hoeffding bound and the Han-Kobayashi bound to the case of a composite alternative hypothesis.
Moreover, we do not need to restrict the alternative hypothesis to i.i.d.\ distributions but can allow permutation invariant or even more general distributions.
We formally state all of our results in Section~\ref{sec:main}.

Our second main result, which is an application of the first, is to give an operational interpretation to various measures of R\'enyi mutual information, R\'enyi conditional entropy, and R\'enyi conditional mutual information.
A complete discussion of this can be found in Section~\ref{sec:operational}, and here we exhibit a few representative examples:
\begin{enumerate}

  \item Let $(X, Y)$ be two random variables governed by a joint probability distribution $P_{XY}$ with marginal $P_X$. We find that the hypothesis testing problem
  \hypoinformalc{ $(X^n,Y^n) \sim P_{XY}^{\times n}$ }{ $X^n \sim P_{X}^{\times n}$ independent of $Y^n$\label{eq:hpoly}} 
  originally proposed by Polyanskiy~\cite[Sec.~II]{polyanskiy13} in the context of channel coding, leads to an operational interpretation of Sibson's~\cite{sibson69} definition (see also~\cite[p.~27]{csiszar95}) of R\'enyi mutual information,
  \begin{align}
    I_s^{\uda}(X\!:\!Y) = \min_{Q_Y} D_s(P_{XY} \|P_X \times Q_Y) \,.
  \end{align}
  A similar hypothesis testing problem where the alternative hypothesis further requires that $X^n$ is uniform leads to an operational interpretation of Arimoto's definition of R\'enyi conditional entropy~\cite{arimoto75}.
  
  \item We further treat the problem of detecting correlations in a collection of random variables. Specifically, consider a null hypothesis that the random random variables $(X_1, X_2, \ldots, X_k)$ are governed by a specific distribution $P_{X_1 X_2 \ldots X_k}$ and compare this to the alternative hypothesis that these random variables are independent, which is a natural formulation from the viewpoint of statistics.
  This can be phrased as the hypothesis testing problem
    \hypoinformal{ $(X_1^n, X_2^n, \ldots, X_k^n) \sim P_{X_1 X_2 \ldots X_k}^{\times n}$ }{ $(X_1^n, X_2^n, \ldots, X_k^n) \sim \big(  Q_{X_1} \times Q_{X_2} \times \ldots \times Q_{X_k} \big)^{\times n}$ for some $Q_{X_i}$\label{eq:hcomp}}
    
  The quantity $D(P_{X_1 X_2 \ldots X_k} \| P_{X_1} \times P_{X_2} \times \ldots \times P_{X_k})$ is a measure of correlations for $k$-partite systems. We show that it attains operational significance as a threshold rate for the above problem.
  We also derive error exponents and strong converse exponents for this problem as long as $R$ is close enough to the threshold rate. These are determined by the quantities
  \begin{align}
     \min_{ Q_{X_1}, Q_{X_2}, \ldots Q_{X_k} } D_s(P_{X_1 X_2 \ldots X_k} \| Q_{X_1} \times Q_{X_2} \times \ldots \times Q_{X_k}) \,.
  \end{align}
  
    Our test depends on the specific distribution $P_{X_1 X_2 \ldots X_k}$, so it is not able to detect arbitrary correlations in these random variables.

  \item Let $P_{XYZ}$ be a joint probability distribution. We find that the hypothesis testing problem
  \hypoinformalc{ $(X^n,Y^n,Z^n) \sim P_{XYZ}^{\times n}$ }{ $(X^n,Y^n,Z^n) \sim Q_{XYZ}^{\times n}$ for some $Q_{XZY}$ that is Markov $X \leftrightarrow Y \leftrightarrow Z$ } 
  yields an operational interpretation for the conditional mutual information, $I(X\!:\!Z|Y)$, as the threshold rate. Moreover, the error exponents are determined by a certain R\'enyi conditional mutual information,
  \begin{align}
      I_s^{\ddda}(X\!:\!Z|Y) = \min_{Q_{XYZ} } D_s(P_{XYZ} \| Q_Y \times Q_{X|Y} \times Q_{Z|Y}) \,.
  \end{align}
  However, this definition of R\'enyi conditional mutual information is by no means the only definition that attains operational significance. If we vary the problem slightly and only consider alternative hypotheses with a fixed marginal $(X^n, Y^n) \sim P_{XY}^{\times n}$ we recover the same threshold rate but different exponents determined by
    \begin{align}
      I_s^{\uuda}(X\!:\!Z|Y) &= \min_{Q_{Z|Y}} D_s(P_{XYZ} \| P_{XY} \times Q_{Z|Y}) \, .
  \end{align}
  Our results thus yield an operational interpretation for this definition of R\'enyi conditional mutual information for all positive $s$, and we provide a closed form of this quantity in~\eqref{eq:cmi-closed-form}. To the best of our knowledge this definition has not appeared in the literature before. From the operational perspective we have chosen here, it is a natural extension of Sibson's definition of R\'enyi mutual information to the conditional setting, and we expect it to have other applications in information theory.

\end{enumerate}

On a technical level, our work introduces the concept of a universal distribution and a universal channel. The purpose of the former is to dominate any i.i.d.\ product (or permutation invariant) distribution in terms of the relative entropy and the R\'{e}nyi relative entropy (cf.~Lemmas~\ref{lm:universal-limit} and~\ref{lm:uni-dist}). More formally, we show that there exists a sequence of distributions $U^n$ on $\cX^n$ such that $D_s(P^n \| Q^{\times n}) \geq D_s(P^n \| U^n) + O(\log n)$ for any $P^n$, $Q$ uniformly for all $s \geq 0$. 
Similarly, the output distribution of the universal channel dominates the output distribution of any memoryless product (or permutation covariant) channel, whenever both channels are given the same input (cf.~Lemma~\ref{lm:uni-channel}). A similar kind of approximation was discussed perviously by
Davisson~\cite{davisson73} (see also Clarke and Barron~\cite{clarke90}) but they only showed the approximation (using a Baysian mixture) for the case $s = 1$ corresponding to the relative entropy.\footnote{Their result was extended in a recent paper~\cite{hayashi15} to the approximation in terms of the R\'{e}nyi relative entropy, even in the continuous case.} 
We also note that the concept of universal decoding was studied in~\cite{feder98}, but in this work the universal decoder can only dominate finitely many decoders whereas our universal states and channels dominate a continuous set of states and channels, respectively.
Therefore, the methods in References~\cite{davisson73,clarke90,feder98} cannot be directly applied to our analysis. 
The universal distribution is the classical analogue of the universal state originally introduced in the quantum setting by one of the authors in~\cite{hayashi09b} and~\cite{hayashi09c}. The latter paper also introduced universal classical-quantum channels (see also~\cite{fawzirenner14} for a fully quantum universal channel). 

Another contribution is the axiomatic approach we have taken to the problem.
We derive a sufficient condition for the hypotheses testing problems
to derive analogues of Yushkevich's bound, Hoeffding's bound and the Han-Kobayashi bound.
Since this approach accepts hypotheses containing non-i.i.d. distributions, we expect it to have wide applicability.

\subsubsection*{Outline} The remainder of the paper is structured as follows. In Section~\ref{sec:framework} we introduce the axiomatic framework for composite hypothesis testing that we build on, and also define our information quantities, the R\'enyi divergences. In Section~\ref{sec:main} we present our main results and treat some generic examples, with the proofs deferred to the later sections. The examples discussed in the introduction, which yield an operational interpretation of various notions of R\'enyi conditional entropy, mutual information, and conditional mutual information, are then treated in detail in Section~\ref{sec:operational}. In fact, Section~\ref{sec:operational} can be understood without referring to Section~\ref{sec:main}. The proofs for the Hoeffding bound are discussed in Section~\ref{sec:hoeffding}, the Han-Kobayashi bound follows in Section~\ref{sec:sc} and the second order analysis of Stein's lemma is found in Section~\ref{sec:second}.  We conclude our work with a discussion and outlook in Section~\ref{sec:conc}. 


\section{A Framework for Composite Hypothesis Testing}
\label{sec:framework}

In this section we introduce a general framework for composite hypothesis that encompasses, but is not restricted to, the examples mentioned in the introduction. 

\subsection{Binary Hypothesis Testing with Composite Alternative Hypothesis}
\label{sec:binaryhypo}

We restrict our attention to finite alphabets. Let $\cX$ be such an alphabet.
The set of probability mass functions (in the following often just called distributions) on $\cX$ is denoted by $\cP(\cX)$ and comprised of positive valued functions on $\cX$ with $\sum_{x \in \cX} P(x) = 1$. For $P \in \cP(\cX)$ and a random variable $X$ on $\cX$, we write $X \sim P$ to denote that $X$ is distributed according to the law $P$. we use $\cX^n$ to denote the $n$-fold Cartesian product of $\cX$ and its elements by vectors $x^n = (x_1, x_2, \ldots, x_n)$. For any $P \in \cP(\cX)$, we use $P^{\times n}$ to denote the identical and independent distribution (i.i.d.) given by $P^{\times n}(x^n) = \prod_{i=1}^n P(x_i)$.

We consider hypothesis testing problems with a composite alternative hypothesis of the following form. 

\begin{definition}
\label{def:hypo}
A sequence of \emph{hypothesis testing problems with composite alternative hypothesis} is determined by a triple $\mathbb{H} = \big(\cX, P, \{\cQ_n\}_{n \in \mathbb{N}} \big)$, comprised of a finite set $\cX$, a
distribution $P \in \cP(\cX)$, and a sequence of sets $\cQ_n \subseteq \cP(\cX^n)$ for all $n \in \mathbb{N}$. This determines a hypothesis testing problem $\mathbb{H}_n$ for each $n \in \mathbb{N}$. Namely, we consider $n$ instances of a random variable $X$ on $\cX$ and the following two hypotheses.
\hypo{$\mathbb{H}_n$}{$X^n \sim P^{\times n}$}{$X^n \sim Q^n$ for some $Q^n \in {\cQ}_n$}
\end{definition}
We will analyze this problem for sequences of sets $\cQ_n$ that satisfy certain axioms (cf.\ Section~\ref{sec:axioms}). 

For convenience we employ the shorthand notation $\cQ \equiv \cQ_1$ and we use $\overline{\cQ}_n$ to denote the convex hull of $\cQ_n$.
Consider probabilistic hypothesis tests, given by a function
$T: \cX \to [0,1]$, where $T(x)$ is the probability of accepting the null hypothesis when observing $x$. We then define the {type-I error} probability and {type-II error} probability, respectively, as follows:
\begin{align}
  \alpha(T;P) := 
\sum_{x \in \cX} P(x)(1-T(x)), 
 \qquad \textrm{and} \qquad
\beta(T;\cQ) := 
\sup_{Q \in \cQ} \
\sum_{x \in \cX} Q(x) T(x) \,. \label{eq:def-beta}
\end{align}

In this work we focus on asymmetric hypothesis testing.
In this context it is convenient to define the quantity 
$\hat{\alpha}(\mu; P\| \cQ)$ as the minimum type-I error probability when the type-II error probability is below a threshold $\mu \geq 0$, i.e.\ we consider the following optimization problem:
\begin{align}
\hat{\alpha}(\mu;P\| \cQ) := 
\min_{T} \big\{   \alpha(T;P) \,\big|\, 
\beta(T;\cQ)  \leq \mu \big\} , \qquad \textrm{for } \mu \in \mathbb{R} \,.
\end{align}
Note that $\mu \mapsto \hat{\alpha}(\mu;P\| \cQ)$ is monotonically decreasing and evaluates to $0$ for $\mu \geq 1$. Note that we always have 
\begin{align}
  \hat{\alpha}(\mu;P\| \cQ) = \hat{\alpha}(\mu;P\| \overline{\cQ}) \,. \label{eq:convex-hull}
\end{align}
Moreover, since the sum over which we take the supremum in~\eqref{eq:def-beta} is linear in $Q$ and the maximum is thus achieved on the boundary.

\subsection{Operational Quantities for Asymmetric Hypothesis Testing}
\label{sec:opquant}

Let us now discuss the main operational quantities that we want to investigate.

\subsubsection{Threshold Rate}

We will study the following properties of asymmetric composite hypothesis tests. The first concept concerns the threshold rate of a sequence of such tests.
\begin{definition}
  Let $\mathbb{H} = \big(\cX, P, \{\cQ_n\}_{n \in \mathbb{N}} \big)$ be a sequence of hypothesis testing problems. We define the \emph{threshold rate} of the sequence $\mathbb{H}$ as
  \begin{align}
    \Rth(\mathbb{H}) := \sup \bigg\{ R \in \mathbb{R} : \limsup_{n \to \infty} \hat{\alpha}\big( \exp(-n R); P^{\times n} \| \cQ_n\big) = 0 \bigg\} \,.  
  \end{align}
  Similarly, we define the \emph{strong converse threshold rate} of the sequence $\mathbb{H}$ as (cf.\ \cite[Def.~4.3.1]{han02} and~\cite{nagaoka07})
  \begin{align}
    \Rth^*(\mathbb{H}) = \inf \bigg\{ R \in \mathbb{R} : \liminf_{n \to \infty} \hat{\alpha}\big( \exp(-n R); P^{\times n} \| \cQ_n\big) = 1 \bigg\} \,.
  \end{align}
\end{definition}
Clearly $\Rth(\mathbb{H}) \leq \Rth^*(\mathbb{H})$ always holds.
Moreover, we note that the threshold rate is always nonnegative because $\hat{\alpha}\big( \exp(-n R); P^{\times n} \| \cQ_n\big)$ vanishes for $R \leq 0$. 
The problems we study in this paper are particularly well-behaved and we will find that the threshold rate and the strong converse threshold rate agree. 

\subsubsection{Error and Strong Converse Exponents}

Moreover, if we choose a rate $R$ below $\Rth(\mathbb{H})$ then we will observe that $\hat{\alpha}\big( \exp(-n R); P^{\times n} \| \cQ_n\big)$ converges exponentially fast to $0$ as $n$ increases. The exponent characterizing this decrease is called the \emph{error exponent} of the sequence with regards to $R$.
\begin{definition}
  Let $\mathbb{H} = \big(\cX, P, \{\cQ_n\}_{n \in \mathbb{N}} \big)$ be a sequence of hypothesis testing problems. For every $R > 0$, the \emph{error exponent} of $\mathbb{H}$ with regards to $R$ is defined as
\begin{align}
  \mathrm{e}_R(\mathbb{H}) := \liminf_{n \to \infty} - \frac{1}{n} \log \hat{\alpha}\big( \exp(-n R); P^{\times n} \| \cQ_n\big) \,,
\end{align}
if this limit exists, or $+ \infty$ otherwise. 
\end{definition}

If we choose a rate $R$ exceeding $\Rth(\mathbb{H})$ then we will instead observe that $\hat{\alpha}\big( \exp(-n R); P^{\times n} \| \cQ_n\big)$ converges exponentially fast to $1$ as $n$ increases. The exponent characterizing this convergence is called the \emph{strong converse exponent} of the sequence with regards to $R$.
\begin{definition}
  Let $\mathbb{H} = \big(\cX, P, \{\cQ_n\}_{n \in \mathbb{N}} \big)$ be a sequence of hypothesis testing problems. For every $R > 0$, the \emph{strong converse exponent} of $\mathbb{H}$ with regards to $R$ is defined as
\begin{align}
  \mathrm{sc}_R(\mathbb{H}) := \limsup_{n \to \infty} - \frac{1}{n} \log \Big( 1- \hat{\alpha}\big( \exp(-n R); P^{\times n} \| \cQ_n\big) \Big) \,,
\end{align}
if this limit exists, or $+ \infty$ otherwise. 
\end{definition}

\subsection{Information Measures for Asymmetric Hypothesis Testing}
\label{sec:renyi}

All our results will be stated as an equivalence between one of the above-mentioned operational quantities and an information measures derived from the R\'enyi divergence. We formally define the R\'enyi divergence~\cite{renyi61} here.

\begin{definition}
Let $P \in \cP(\cX)$ and $Q : \cX \to [0, \infty)$. For $s \in (0,1) \cup (1,\infty)$, define
\begin{align}
 \label{eq:renyi-g}
   g_s(P\|Q) := \lim_{\varepsilon \to 0}\ \sum_{x \in \cX} P(x)^{s} \big( Q(x) + \varepsilon \big)^{1-s} = \sum_{\substack{x \in \cX \\ P(x) > 0}} P(x)^{s} Q(x)^{1-s} ,
\end{align}
   where the latter expression is finite either if $s < 1$ or if $Q(x) = 0$ implies $P(x) = 0$ for all $x \in \cX$. Otherwise, we set $g_s(P\|Q) = +\infty$.
   The R\'enyi divergence of $P$ with regards to $Q$ of order $s$ is then defined as
\begin{align}\label{eq:renyi}
  D_{s}(P \| Q) &:=  \frac{\log g_{s}(P \| Q)}{s-1}
\end{align}
For $s \in \{0,1,\infty\}$ the R\'enyi divergence is defined as the corresponding limit.
\end{definition}
See~\cite{vanerven14} for a comprehensive discussion of its properties, which we summarize here. The R\'enyi divergence is lower semi-continuous and diverges to $+\infty$ if the support condition is violated and $s > 1$ or if $s < 1$ and $P$ and $Q$ do not share any support, in which case $g_s(P\|Q) = 0$. Our sets $\cQ_n$ may contain elements that violate these conditions. Nonetheless, sets consistent with our axioms will always contain at least one element that satisfies the support conditions.
One of the most important properties of the latter functional is that $g_{s}(\cdot\|\cdot)$ is jointly concave for $s \in (0,1)$ and jointly convex for $s \in (1,\infty)$. Moreover, the function $s \mapsto \log g_{s}(P\|Q)$ is convex and $s \mapsto D_{s}(P\|Q)$ is monotonically increasing.
Furthermore, the Kullback-Leibler divergence is given by
\begin{align}
  D(P\|Q) := D_1(P\|Q) = \sum_x P(x) \big( \log P(x) - \log Q(x) \big) \,.\label{eq:kullback}
\end{align}
For two positive valued functions $Q$ and $Q'$ on $\cX$, we observe that $Q(x) \leq Q'(x)$ for all $x \in \cX$ implies $D_{s}(P\|Q) \geq D_{s}(P\|Q')$. Furthermore, for any scalar $v$, we have $D_{s}(P\|vQ) = D_{s}(P\|Q) - \log v$.

To present the second order of Stein's lemma we need to introduce some additional quantities. The \emph{variance of the logarithmic likelihood ratio} is given by
\begin{align}
  V(P\|Q) := \sum_{x \in \cX} P(x) \big( \log P(x) - \log Q(x) - D(P\|Q) \big)^2 \,. \label{eq:optimal-crit}
\end{align}
 This variance is proportional to the derivative of the R\'enyi divergence at $s = 1$, a consequence of the fact that the R\'enyi divergence is proportional to the cumulant generating function of the logarithmic likelihood ratio.
  More precisely, the first order Taylor expansion of $D_s(P\|Q)$ around $s = 1$ is given by
  \begin{align}
    D_s(P\|Q) = D(P\|Q) + \frac{s-1}{2 \log e} V(P\|Q) + O(s^2) \label{eq:taylor} \,.
  \end{align}

Finally, we define the R\'enyi divergence of $P$ with regards to a set $\cQ$ of positive valued functions as
\begin{align}
  D_{s}(P \| \cQ) := \inf_{Q \in \cQ} D_{s}(P \| Q) \,. \label{eq:minimizers}
\end{align}
The minimizer, if it is unique, is defined as 
\begin{align}
  \hat{Q}^{s} := \argmin_{Q \in \cQ} D_{s}(P\|Q) \,. \label{eq:mini2}
\end{align}
We define $V(P\|\cQ) := V(P\|\hat{Q}^1)$.
Similarly, taking note of the sign of $(s-1$), we define
\begin{align}
  g_{s}(P\|\cQ) := \exp\big( (s-1) D_{s}(P\|\cQ) \big) = \begin{cases} \displaystyle \sup_{Q \in \cQ} g_{s}(P \| Q) & \textrm{if } s \in (0,1) \\
  \displaystyle \inf_{Q \in \cQ} g_{s}(P \| Q) & \textrm{if } s \in (1,\infty)
  \end{cases}.
\end{align}

\subsection{Axioms for Alternative Hypotheses}
\label{sec:axioms}

Let us fix a probability distribution $P \in \cP(X)$. We present a collection of axioms that the sets $\{ \cQ_n \}_{n \in \mathbb{N}}$ must satisfy in order for our main results to hold. 
The first axiom ensures that the base set, $\cQ$, is convex. 
\begin{axiom}[convexity]
  \label{ax:convex}
  The set $\cQ \subseteq \cP(\cX)$ is compact convex. Moreover, for all $s > 0$ the minimizer $\hat{Q}^s$ in~\eqref{eq:mini2}
  is unique and lies in $\relint(\cQ)$.\footnote{The relative interior of a convex set $\Theta$ is the set $\relint(\Theta) := \{ x \in \Theta : \forall y \in \Theta \,\exists \lambda > 1 \textrm{ s.t. } \lambda x + (1-\lambda) y \in \Theta \}$.}
\end{axiom}

The second axiom ensures that i.i.d.\ products of distributions in $\cQ$ are elements of $\cQ_n$.

\begin{axiom}[product distributions]
  \label{ax:prod}
  The set $\cQ_n$ contains the element $Q^{\times n}$ for every $Q \in \cQ$.
\end{axiom}

As a direct consequence of Axiom~\ref{ax:prod} and the additivity of the R\'enyi divergence for product distributions, we find the following lemma.
\begin{lemma}[subadditivity]
  \label{lm:subadd}
  Assuming Axiom~\ref{ax:prod}, we have $D_{s}(P^{\times n} \| \cQ_n) \leq n D_{s}(P\|\cQ)$ for all $s > 0$ and $n \in \mathbb{N}$.
\end{lemma}
The purpose of the next axiom is to ensure that this is in fact an equality.

\begin{axiom}[additivity] \label{ax:add}
  For all $s > 0$ and $n \in \mathbb{N}$, we have $D_{s}(P^{\times n} \| \cQ_n) \geq n D_{s}(P\|\cQ)$.
\end{axiom}

The next axiom concerns the existence of a sequence of universal distributions. Before we state it, let us introduce some additional notation. We denote the {symmetric group} of permutations of $n$ elements by $S_n$. This group has a natural representation as bistochastic matrices. For every $\pi \in S_n$, the matrix $W^n[\pi]$ is defined by the relation $P^n W^n[\pi] (x_1, x_2, \ldots, x_n) = P^n (x_{\pi(1)}, x_{\pi(2)}, \ldots, x_{\pi(n)})$. We say that a probability distribution $P^n \in \cP(\cX^n)$ is {permutation invariant} if $P^n W^n[\pi] = P^n$ holds for all $\pi \in S_n$. The set of all permutation invariant distributions on $\cX^n$ is denoted $\cP^{\sym}(\cX^n)$ and permutation invariant distributions in $\cQ_n$ comprise the subset $\cQ_n^{\sym}$.

\begin{axiom}[universal distribution]
  \label{ax:universal}
  There exists a sequence of probability mass functions $\{ U^n \}_{n \in \mathbb{N}}$ with $U^n \in \cP^{\sym}(\cX^n)$ and a polynomial $v(n)$ such that the following relation holds. For all $n \in \mathbb{N}$ and $Q \in \cQ_n^{\sym}$, 
  \begin{align}
    Q(x^n) \leq v(n)\, U^n(x^n), \quad  
     \forall \, x^n \in \cX^n, \qquad \textrm{and} \qquad  D_{s}(P^{\times n}\| U^n) \geq D_{s}(P^{\times n}\| \cQ_n) \,. \label{eq:ax-uni}
  \end{align}
The former condition is rewritten as
$D_{\max}( Q \|  U^n) \le \log v(n)$ for $Q \in \cQ_n^{\sym}$,
where $D_{\max}( Q \|  P):=  \log \max_{x} \frac{Q(x) }{P(x)}$.
Moreover, the set $\cQ_n$ is closed under symmetrization, i.e.\ if $P^n \in \cQ_n$ then $\frac{1}{n!} \sum_{\pi \in S_n} P^n W^n[\pi] \in \cQ_n^{\sym}$.
\end{axiom}

The latter condition in~\eqref{eq:ax-uni} is automatically satisfied if $U^n \in \cQ_n^{\sym}$, but this is not necessary.
The above axioms have the following immediate consequence.

\begin{lemma}
\label{lm:universal-limit}
  Assume Axioms~\ref{ax:add} and~\ref{ax:universal} hold. Then, for all $s > 0$
  \begin{align}
    \lim_{n \to \infty} \frac{1}{n} D_{s}(P^{\times n} \| U^n) = D_{s}(P \| \cQ) \,, \label{eq:uni1}
  \end{align}
   and the map $s \mapsto \phi(s) := \log g_s(P\|\cQ)$ is convex. 
\end{lemma}

\begin{IEEEproof}
  We first show~\eqref{eq:uni1}.
  Additivity implies that $\frac{1}{n} D_{s}(P^{\times n}\| U^n) \geq \frac{1}{n} D_{s}(P^{\times n}\| \cQ_n) = D_{s}(P\|\cQ)$. To establish the other direction, we use Axiom~\ref{ax:universal}. For any $Q \in \cQ$, we have
  \begin{align}
    \frac{1}{n} D_{s}(P^{\times n}\| U^n) \leq 
    \frac{1}{n} \Big( D_{s}(P^{\times n} \| Q^{\times n} ) + \log v(n) \Big)
    = D_{s}(P\|Q) + \frac{1}{n} \log v(n) \label{eq:step1}
  \end{align}
  Hence, minimizing over all such $Q$ we can replace $D_{s}(P\|Q)$ with $D_{s}(P\|\cQ)$ on the right hand side.
  Moreover, using the property that $v(n)$ grows polynomially in $n$, we find $\limsup_{n\to\infty} \frac{1}{n} D_{s}(P^{\times n}\| U^n) \leq D_{s}(P\|\cQ)$.
  Finally, since $s \mapsto \log g_s(P\|\cQ)$ is the point-wise limit of convex functions, it is also convex.
\end{IEEEproof}

Finally, we note that convexity in Axiom~\ref{ax:convex} is quite a strong requirement and not satisfied by some of the examples we consider. Instead of requiring that the set $\cQ$ is convex, it suffices to assume that there exists a convenient convex parametrization of the set such that concavity and convexity of $g_{s}(P\|\cdot)$ are preserved. 

\begin{axiom}[convex parametrization, replaces Axiom~\ref{ax:convex}]\label{ax:para}
  There exists a compact convex set $\Theta$ in a finite-dimensional vector space, a twice continuously differentiable ($C^2$) function $\Theta \ni \theta \mapsto Q_{\theta} \in \cQ$, and an open interval $(a,b)$ containing $1$ such that the following holds:
  \begin{itemize}
    \item We have $\cQ = \{ Q_{\theta} : \theta \in \Theta \}$.
    \item For all $s \in (a,b)$, the minimizer $\hat{\theta}^{s} := \argmin_{\theta \in \Theta} D_{s}(P\|Q_{\theta})$ is unique and lies in $\relint(\Theta)$.
    \item The map $\theta \mapsto g_{s}(P\|Q_{\theta})$ has negative definite Hessian at $\hat{\theta}^{s}$ for all $s \in (a,1)$ and positive definite Hessian at $\hat{\theta}^{s}$ for all $s \in (1,b)$. Moreover, the map $\theta \mapsto D_{s}(P\|Q_{\theta})$ has positive definite Hessian at $\hat{\theta}^{1}$.
  \end{itemize}
\end{axiom} 
Since the map in Axiom~\ref{ax:para} is assumed to be $C^2$ and $(s, Q) \mapsto D_{s}(P\|Q)$ is $C^2$, Axiom~\ref{ax:para} implies that $(s,\theta) \mapsto D_{s}(P\|Q_{\theta})$ is $C^2$ as well.
Further note that Axiom~\ref{ax:convex} implies Axiom~\ref{ax:para} using the trivial parametrization $\Theta = \cQ$ and $(a,b) = (0,\infty)$.\footnote{However, the converse argument is not true because for any two points $\theta$ and $\theta'$ and $\lambda \in (0,1)$, the distribution  $Q_{\lambda \theta+ (1-\lambda) \theta'} \in {\cal Q}$ does not necessarily equal the distribution $\lambda Q_{\theta}+ (1-\lambda) Q_{\theta'} \in \cP(\cX)$. That is, the convex combination in the parameter space is different from the convex combination in $\cP(\cX)$, in general. (Examples will be given in Section~\ref{sec:examples}.)} If we assume Axiom~\ref{ax:para} instead of Axiom~\ref{ax:convex} we must also relax the additivity property. Namely, additivity in Axiom~\ref{ax:add} is only required in the interval $s \in (a,b)$ and Lemma~\ref{lm:universal-limit} only holds for $s \in (a,b)$.

\section{Main Results and Examples}
\label{sec:main}

\subsection{Statement of Main Results}

Our first result considers the asymptotic situation where the
type-II error probability goes to zero exponentially with a rate below $D(P\|\cQ)$. 
In this case, we find that type-I error probability converges to zero exponentially fast, 
and the exponent is determined by the R\'enyi divergence, $D_{s}(P\|\cQ)$ with $s < 1$. 

To state our result we need the following concept.
\begin{definition}
\label{def:critical}
  Fix $P \in \cP(\cX)$ and $\cQ \subseteq \cP(\cX)$. For any $c \geq 0$, the $c$-critical rate is defined as 
  \begin{align}
    R_c := \lim_{s \to c} \big( s\phi'(s) - \phi(s) \big)
  \end{align}
  with $\phi(s) = (s-1) D_s(P\|\cQ)$ as defined in Lemma~\ref{lm:universal-limit}.
\end{definition}
The map $c \mapsto R_c$ on $(a, b)$ is monotonically increasing  (cf.\ Lemma~\ref{lm:hmono}), and furthermore we find $R_0 = D_0(P\|\cQ)$ and $R_1 = D(P\|\cQ)$, as well as $R_{\infty} \geq D_{\infty}(P\|\cQ)$.

\begin{theorem}\label{th:hoeffding}
Let $\mathbb{H} = \big(\cX, P, \{\cQ_n\}_{n \in \mathbb{N}} \big)$ be such that Axioms~\ref{ax:prod}--\ref{ax:para} are satisfied on $(a,1]$.
Then, for any $R > R_a$,
  \begin{equation}
  \label{eq:hoeffding-thm}
  \mathrm{e}_R(\mathbb{H}) 
   = \sup_{s \in (a, 1)} \left\{ \frac{1-s}{s} \big( D_{s}(P\|\cQ)   - R \big) \right\}.
  \end{equation}
\end{theorem}
The proof is given in Section~\ref{sec:hoeffding}. The case where $\cQ_n = \{ Q^{\times n} \}$ are singletons is attributed to Hoeffding~\cite{hoeffding65}. 
Note that if $R \geq D(P\|\cQ)$ the right hand side of~\eqref{eq:hoeffding-thm} evaluates to zero, revealing that in this case the error of the first kind will decay slower than exponential in $n$. Otherwise the right hand side is always positive. 

Our second result considers the case where type-II error probability goes to zero exponentially
with a rate exceeding the mutual information $D(P\|\cQ)$. In this case, we find that type-I error
probability converges to 1 exponentially fast, and the exponent is determined by the
R\'enyi divergence $D_s(P\|\cQ)$, with $s > 1$.

\begin{theorem}\label{th:sc}
Let $\mathbb{H} = \big(\cX, P, \{\cQ_n\}_{n \in \mathbb{N}} \big)$ be such that Axioms~\ref{ax:prod}--\ref{ax:para} are satisfied on $[1, b)$.
Then, for any $R < R_b$,
  \begin{equation}
  \label{eq:sc-thm}
    \mathrm{sc}_R(\mathbb{H}) = \sup_{s \in (1, b)} \left\{ \frac{s-1}{s} \big( R - D_{s}(P\|\cQ) \big) \right\}.
  \end{equation}
\end{theorem}
The proof is given in Section~\ref{sec:sc}. The case where $\cQ_n$ are singletons is attributed to Csisz\'ar-Longo~\cite{csiszar71} and Han-Kobayashi~\cite{han89}. Note that even in the singleton case, the original results do not apply for $R > R_{\infty}$. In fact Nakagawa-Kanaya~\cite{nakagawa93} showed that in this setting the above optimal exponent can be attained only by a randomized test. We will not further discuss this case.

Again we note that if $R \leq D(P\|\cQ)$ the right hand side of~\eqref{eq:sc-thm} evaluates to zero, otherwise it is strictly positive. 
The threshold rates are thus determined by the above results, and we find the following corollary of Theorems~\ref{th:hoeffding}~and~\ref{th:sc}.

\begin{corollary}
  Let $\mathbb{H} = \big(\cX, P, \{\cQ_n\}_{n \in \mathbb{N}} \big)$ be such that Axioms~\ref{ax:prod}--\ref{ax:para} are satisfied with any $(a,b)$ containing $1$. Then,
  \begin{align}
    \Rth(\mathbb{H}) = \Rth^*(\mathbb{H}) = D(P\|\cQ) \,.
  \end{align}
\end{corollary}

For completeness, we also investigate the second order behavior, namely we investigate the error of the first kind when the error of the second kind vanishes as $\exp(- n D(P \|\cQ ) - \sqrt{n}r)$.
This analysis takes a step beyond Stein's lemma and extends Yushkevich's work for simple alternative hypotheses~\cite{yushkevich53}. 
\begin{theorem}
\label{th:second}
  Assume Axioms~\ref{ax:prod}--\ref{ax:para} hold for any $(a,b) \ni \{1\}$. Then, for any $r \in \mathbb{R}$, we have
  \begin{align}
    \lim_{n \to \infty}  \hat{\alpha}\Big( \exp\big(- n D(P \|\cQ ) - \sqrt{n}r\big) ; P^{\times n} \Big\| \cQ_n \Big)  = \Phi \left( \frac{r}{\sqrt{V(P\|\cQ)}} \right) ,
  \end{align}
  where $\Phi(x) := (2\pi)^{-\frac12} \int_{-\infty}^{x} e^{-\frac{y^2}{2}} \mathrm{d}y$ and $V(P\|\cQ)$ is defined in the line following~\eqref{eq:mini2}.
\end{theorem}

The proof is given in Section~\ref{sec:second}. The achievability proof is of a different flavor than previous proofs of the singleton case and relies on L\'evy's continuity theorem.

\subsection{Examples}
\label{sec:examples}

In the following we will discuss various examples of hypothesis testing problems that can be tackled with the above framework. The cases we treat here in particular cover the examples in Section~\ref{sec:operational}. 

\subsubsection{Product distributions with a fixed marginal}
\label{sec:ex1}

Let $\cX$ and $\cY$ be two finite sets. Consider a pair of random variables $(X, Y) \sim P_{XY}$ that are governed by a joint probability distribution $P_{XY} \in \cP(\cX \times \cY)$. We denote by $P_{X|Y=y}$ the distribution of $X$ conditioned on the event $Y=y$. The conditional distribution $P_{X|Y}$ is interpreted as a stochastic matrix mapping or channel from $\cY$ to $\cX$. In particular, we write $P_{XY} = P_Y \times P_{X|Y}$ and $P_X = P_Y P_{X|Y}$. 
We assume without loss of generality that $P_X$ and $P_Y$ have full support, i.e.\ we restrict the sets $\cX$ and $\cY$ to the support of the marginals $P_X$ and $P_Y$, respectively.
Moreover, let $T_X \in \cP(\cX)$ be such that $T_X$ has full support as well. 
Now consider the sets
\begin{align}
  \cQ_n = \big\{ T_X^{\times n} \times Q_{Y^n} :\ Q_{Y^n} \in \cP(\cY^n) \big\} \,. \label{eq:set-one}
\end{align}
We emphasize that the distributions $Q_{Y^n}$ are unstructured, in particular they do not have to be $n$-fold i.i.d.\ products.


\begin{prop}
  \label{pr:examples-one}
  The sequence of tests $\big( \cX \times \cY, P_{XY}, \{\cQ_n \}_{n \in \mathbb{N}} \big)$ with $\cQ_n$ in~\eqref{eq:set-one} satisfies Axioms~\ref{ax:convex}--\ref{ax:universal}. Moreover, this still holds if we restrict $\cQ_n$ to permutation invariant or i.i.d.\ product distributions.
\end{prop}

The proof is given in Appendix~\ref{sec:proof-examples-1} and relies on the following Lemma, which might be of independent interest.

\begin{lemma}
  \label{lm:uni-dist}
  The exists a sequence of distributions $\{ U_{X^n}^n \}_{n \in \mathbb{N}}$ with $U_{X^n}^n \in \cP^{\sym}(\cX^n)$ such that, for every $n \in \mathbb{N}$ and $Q_{X^n} \in \cP^{\sym}(\cX^n)$, we have  $Q_{X^n}(x^n) \leq |\cT_n(\cX)|\, U_{X^n}^n(x^n)$ for all $x^n \in \cX^n$.
\end{lemma}

In the above lemma we used $\cT_n(\cX)$ to denote the set of $\cX$-types of length $n$.
This is the classical analogue of the universal state originally introduced in the quantum setting~\cite{hayashi09b,hayashi09c}. 
The proof uses the method of types~\cite{csiszar98}, and the result is only sensible for finite sets.

\begin{IEEEproof}
The universal distributions are given by
  \begin{align}
    U_{X^n}^n(x^n) = \sum_{\lambda \in \cT_n(\cX)} \frac{1}{| \cT_n(\cX) |} \frac{ 1 }{ |\lambda| } \, 1 \{ x^n \textrm{ is of type } \lambda \} \, ,
  \end{align}
  where $\frac{1}{|\lambda|} 1 \{ x^n \textrm{ is of type } \lambda \}$ is the uniform distribution over all sequences of type $\lambda$. Every permutation invariant distribution in $Q_{X^n} \in \cP^{\sym}(\cX)$ has to be flat over sequences of the same type. Namely, it has to be of the form
  \begin{align}
    Q_{X^n}(x^n) = \sum_{\lambda \in \cT_n(\cX)}  \frac{ q(\lambda) }{ |\lambda| } \, 1 \{ x^n \textrm{ is of type } \lambda \} \label{eq:universal1} \,
  \end{align}
  for some distribution $q \in \cP(\cT_n(\cX))$. The desired bound can now be verified easily.
\end{IEEEproof}

\subsubsection{General (permutation invariant) product distributions}
\label{sec:ex2}

Consider finite sets $\cX_1$, $\cX_2$, \ldots, $\cX_k$ and a distribution $P_{X_1X_2\ldots X_k Y} \in \cP(\cX_1 \times \cX_2 \times \ldots \times \cX_k \times \cY)$. Without loss of generality we assume that all the marginals of $P_{X_1X_2\ldots X_k Y}$ have full support.
Then, consider
\begin{align}
  \cQ_n = \big\{ Q_{X_1^n} \times Q_{X_2^n} \times \ldots \times Q_{X_k^n} \times Q_{Y^n} :\  Q_{X_i^n} \in \cP^{\sym}(\cX_i^n), i \in [k] \textrm{ and } Q_{Y^n} \in \cP(\cY^n) \big\} , \label{eq:set-two} \,.
\end{align}
Note that these sets are not convex, so Axiom~\ref{ax:convex} is certainly violated. Moreover, note that the  restriction that the $Q_{X_i^n}$ be permutation invariant is necessary. Without such a restriction, even a correlated null hypothesis lies in the convex hull of the set of alternative hypothesis since
$\cP(\cX \times \cY)$ equals the convex hull of $\cP(\cX) \times \cP(\cY)$. Clearly it is then no longer possible to distinguish these two hypotheses.

\begin{prop}
  \label{pr:examples-two}
  The sequence of tests $\big( \cX_1 \times \cX_2 \times \ldots \times \cX_k \times \cY, P_{X_1X_2\ldots X_k Y}, \{\cQ_n \}_{n \in \mathbb{N}} \big)$ with $\cQ_n$ in~\eqref{eq:set-two} satisfy Axioms~\ref{ax:prod}--\ref{ax:para} with $(a,b) = \big(\frac{k}{k+1}, \infty\big)$. Moreover, this still holds if we restrict $\cQ_n$ to i.i.d.\ product distributions.
\end{prop}

The proof is given in Appendix~\ref{sec:proof-examples-2}.

\subsubsection{Recovered and other Markov distributions}
\label{sec:ex3}

Let $P_{XYZ} \in \cP(\cX \times \cY \times \cZ)$ be a joint probability distribution with marginals $P_X$ and $P_Y$ and $P_Z$. A natural test considers alternative hypothesis comprised of Markov distributions where only two marginals are fixed. We can see this as the problem of distinguishing a fixed tripartite distribution $P_{XYZ}$ from the set of distributions that can be ``recovered'' from its marginal $P_{XY}$ via a probabilistic operation.

 We assume without loss of generality that $P_X(x)$ and $P_Y(y)$ and $P_Z(z)$ have full support, i.e.\ we restrict the sets $\cX$, $\cY$, $\cZ$ to the support of the marginals $P_X$, $P_Y$ and $P_Z$, respectively. The set of conditional probability distributions of $Z$ given $Y$, or channels from $\cY$ to $\cZ$, is denoted by $\cP(\cZ|\cY)$.
Consider the sets
\begin{align}
  \cQ_n = \big\{ P_{XY}^{\times n} \times Q_{Z^n|Y^n} : Q_{Z^n|Y^n} \in \cP(\cZ^n|\cY^n) \big\} \,. \label{eq:set-three}
\end{align}

\begin{prop}
  \label{pr:examples-three}
  The sequence of tests $\big( \cX \times \cY \times \cZ, P_{XYZ}, \{\cQ_n \}_{n \in \mathbb{N}} \big)$ with $\cQ_n$ in~\eqref{eq:set-three} satisfy Axioms~\ref{ax:convex}--\ref{ax:universal}. Moreover, this still holds if we restrict $\cQ_n$ to permutation invariant or i.i.d.\ product distributions.
\end{prop}

This proposition relies on the following lemma, which is of independent interest.

\begin{lemma}
  \label{lm:uni-channel}
  There exists a sequence of channels, $\{ U^n_{Y^n|X^n} \}_{n \in \mathbb{N}}$, where $U^n_{Y^n|X^n} \in \cP^{\sym}(\cY^n|\cX^n)$ such that the following holds. For every $n \in \mathbb{N}$, $Q_{Y^n|X^n} \in \cP^{\sym}(\cY^n|\cX^n)$ and $P_{X^n} \in \cP(\cX^n)$, we have
  \begin{align}
    P_{X^n} \times Q_{Y^n|X^n} (x^n, y^n) \leq | \cT_n(\cX\times\cY) | \, P_{X^n} \times U^n_{Y^n|X^n} (x^n, y^n), \qquad \forall x^n \in \cX^n, y^n \in \cY^n \,.
  \end{align}
\end{lemma}

Both of the above statements, Proposition~\ref{pr:examples-three} and Lemma~\ref{lm:uni-channel}, are proven in Appendix~\ref{sec:proof-examples-3}.

Another natural question is to distinguish between a null hypothesis $P_{XYZ}$ and all Markov distributions $X \leftrightarrow Y \leftrightarrow Z$, i.e.\ distributions $Q_{XYZ} = Q_Y \times Q_{X|Y} \times Q_{Z|Y}$.
Consider the set
\begin{align}
  \cQ_n = \big\{  Q_{X^nY^nZ^n} \in \cP^{\sym}\big((\cX\!\times\!\cY\!\times\!\cZ)^n\big) : Q_{X^nY^nZ^n} = Q_{Y^n} \times Q_{X^n|Y^n} \times Q_{Z^n|Y^n} \big\} \,. \label{eq:set-four}
\end{align}

\begin{prop}
  \label{pr:examples-four}
  The sequence of tests $\big( \cX \times \cY \times \cZ, P_{XYZ}, \{\cQ_n \}_{n \in \mathbb{N}} \big)$ with $\cQ_n$ in~\eqref{eq:set-four} satisfy Axioms~\ref{ax:prod}--\ref{ax:para} with $(a,b) = (\frac23,\infty)$. Moreover, this still holds if we restrict $\cQ_n$ to i.i.d.\ product distributions.
\end{prop}
Again note that every distribution is contained in the convex hull of all Markov distributions, and hence some restrictions on the set are necessary. It is possible to slightly weaken the condition that $(X^n,Y^n,Z^n)$ is permutation invariant, but we will not discuss this here. 
This is shown in Appendix~\ref{sec:proof-examples-4}.


\section{Operational Interpretation of R\'enyi Information Measures}
\label{sec:operational}

In this section we present the main application of our results, finding operational interpretations of various measures of R\'enyi mutual information, conditional entropy and conditional mutual information.

\subsection{R\'enyi Mutual Information: Testing Against Independent Distributions}
\label{sec:examples-mi}

It is well known that the mutual information can be expressed in terms of the Kullback-Leibler divergence in several ways. Consider two random variables $X$ and $Y$ and a joint distribution $P_{XY} \in \cP(\cX \times \cY)$ with marginals $P_X$ and $P_Y$. We are interested in the identities
\begin{align}
  I(X \!:\! Y) &= D(P_{XY} \| P_X \times P_Y)  \label{eq:mi1} \\
          &= \min_{Q_Y \in \cP(Y)} D(P_{XY} \| P_X \times Q_Y) \label{eq:mi2} \\
          &= \min_{Q_X \in \cP(X),\, Q_Y \in \cP(Y)} D(P_{XY} \| Q_X \times Q_Y) \label{eq:mi3}
\end{align}

Each of these identities gives rise to a different hypothesis testing problem and a different notion of R\'enyi mutual information. In the following we treat these three problems in the above order.

\subsubsection{All marginals fixed}
\label{sec:examples-mi-zero}

As a warmup consider the following (simple) hypothesis testing problem. 
\hypo{$\mathbb{H}^{\rm mi}_{\uua}$}{$(X^n, Y^n) \sim P_{XY}^{\times n}$}{$X^n \sim P_X^{\times n}$ and $Y^n \sim P_Y^{\times n}$ are independent}
That is, we set $\cQ_n = \{ P_X^{\times n} \times P_Y^{\times n} \}$ for all $n$ in Defintion~\ref{def:hypo}.  Stein's Lemma and its strong converse ensure that
\begin{align}
  \Rth(\mathbb{H}^{\rm mi}_{\uua}) = \Rth^*(\mathbb{H}^{\rm mi}_{\uua}) 
  = D(P_{XY} \| P_X \times P_Y) = I(X\!:\!Y) \,.
\end{align}
Moreover, the Hoeffding~\cite{hoeffding65} and Han-Kobyashi~\cite{han89} bounds give an operational interpretation for the following R\'enyi mutual information:
\begin{align}
  I_s^{\uua}(X\!:\!Y) := D_s(P_{XY} \| P_X \times P_Y) 
  = \frac{1}{s - 1} \log \left(  \sum_{y \in \cY} P_Y(y) \sum_{x \in \cX}  P_{X|Y=y}(x)^s P_X(x)^{1-s} \right) \,.
\end{align}
As an example, in wire-tap channel coding, this kind of R\'enyi mutual information is used to express the exponents of leaked mutual information~\cite{hayashi11b,han14}.

\subsubsection{One marginal fixed}
\label{sec:examples-mi-one}

Here we consider a hypothesis test where the alternative hypothesis is comprised of product distributions where one marginal is fixed.
This is the example discussed in Section~\ref{sec:ex1} with $T_X = P_X$. 

This hypothesis test figures prominently when analyzing the converse to various channel coding questions in the fixed error regime, for example for second-order analysis of the discrete memoryless channels~\cite{polyanskiy10,hayashi09} and beyond~\cite{tomamicheltan12}. In this context, Polyanskiy~\cite[Sec.~II]{polyanskiy13} raised the following hypothesis testing problem:
\hypoinformal{$(X^n, Y^n) \sim P_{XY}^{\times n}$}{$X^n \sim P_X^{\times n}$ independent of $Y^n$\label{eq:hchannel}}

This problem has the same threshold, $I(X\!:\!Y)$, but gives an operational interpretation for
Sibson's~\cite{sibson69} definition of R\'enyi mutual information, which is given by
\begin{align}
  I_s^{\uda}(X\!:\!Y) &:= \min_{Q_Y \in \cP(\cY)} D_s(P_{XY} \| P_X \times Q_Y) \\
  &\,= \frac{s}{s-1} \log \left( \sum_{y \in \cY} P_Y(y) \Bigg( \sum_{x \in \cX} P_{X|Y=y}(x)^s P_{X}(x)^{1-s} \Bigg)^{\frac{1}{s}} \right) \\
  &\,= \frac{s}{s-1} E_0\left( \frac{s-1}{s} , P_X\! \right) ,
\end{align}
where $E_0$ is Gallager's error exponent function~\cite{gallager68}.\footnote{Verd\'u~\cite{verdu15} recently surveyed Sibson's definition and pointed out its favorable mathematical properties in the case of general alphabets.} The explicit form of the distribution $Q_{Y}$ that achieves the minimum is given by Sibson's identity (cf.~Appendix~\ref{sec:proof-examples}). 

Our results for the optimal error and strong converse exponents then read
\begin{align}
  \mathrm{e}_R(\mathbb{H}^{\rm mi}_{\uda}) = \sup_{s \in (0,1)} \frac{1-s}{s} \big( I_s^{\uda}(X\!:\!Y) - R \big) 
  \quad \textrm{and} \quad
  \mathrm{sc}_R(\mathbb{H}^{\rm mi}_{\uda}) = \sup_{s > 1} \frac{s-1}{s} \big( R - I_s^{\uda}(X\!:\!Y) \big) \,,
\end{align}
In the setting of channel coding with constant composition codes of type $P_X$, the exponents
$\mathrm{e}_R(\mathbb{H}^{\rm mi}_{\uda})$ and 
$ \mathrm{sc}_R(\mathbb{H}^{\rm mi}_{\uda})$
are equal to the error exponent~\cite{gallager68} and the strong converse exponents~\cite{arimoto73}, respectively.
In wire-tap channel coding, it is used for expressing the exponents of leaked information when the leaked information is measured in terms of the variational distance~\cite[Thm.~5]{hayashi13}.

\subsubsection{Arbitrary product distributions, permutation invariant}
\label{sec:examples-mi-two}

Let us now consider the most general problem, 
\hypo{$\mathbb{H}^{\rm mi}_{\dda}$}{$(X^n, Y^n) \sim P_{XY}^{\times n}$}{$X^n$ and $Y^n$ independent, $(X^n, Y^n)$ permutation invariant}
This is covered by the example in Section~\ref{sec:ex2}.
In fact, it is sufficient to require that either $X^n$ or $Y^n$ be permutation invariant (cf.~Proposition~\ref{pr:examples-two}).

We find that for $R$ sufficiently close to the threshold $I(X\!:\!Y)$, we have
\begin{align}
  \mathrm{e}_R(\mathbb{H}^{\rm mi}_{\dda}) = \sup_{s \in (\frac12,1)} \frac{1-s}{s} \big( I_s^{\dda}(X\!:\!Y) - R \big) 
  \quad \textrm{and} \quad
  \mathrm{sc}_R(\mathbb{H}^{\rm mi}_{\dda}) = \sup_{s > 1} \frac{s-1}{s} \big( R - I_s^{\dda}(X\!:\!Y) \big) \,,
\end{align}
with a different definition of R\'enyi mutual information,
\begin{align}
  I_s^{\dda}(X\!:\!Y) := \min_{Q_X \in \cP(\cX),\, Q_Y \in \cP(\cY)} D_s(P_{XY} \| Q_X \times Q_Y) \,.
\end{align}
Our result gives an operational interpretation for this definition of R\'enyi mutual information. However, this operational interpretation only applies for $s \geq \frac12$. In fact, it is unclear if our results can be extended to smaller $s$ and we do not know of a closed form expression for this quantity. While this work was completed, the properties of this definition have been independently studied in~\cite{pfisterlapidoth16}.   

\subsection{R\'enyi Conditional Entropy: Testing Against Uniform and Independent Distribution}
\label{sec:examples-ce}

The conditional entropy can be expressed in terms of the Kullback-Leiber divergence very similarly to the mutual information, with the difference that we require one marginal to be fixed to a uniform distribution. This leads to the following two expressions:
\begin{align}
  H(X|Y) &= \log |\cX| - D( P_{XY} \| R_X \times P_Y) \label{eq:c1} \\
  &= \log |\cX| - \min_{Q_Y \in \cP(\cY)} D( P_{XY} \| R_X \times Q_Y),  \label{eq:c2}
\end{align}
where $R_X$ is the uniform distribution over $X$ and, as in the previous section, $X$ and $Y$ are two random variables governed by a joint distribution $P_{XY} \in \cP(\cX \times \cY)$ with marginals $P_X$ and $P_Y$.
Note that the term $\log |\cX|$ can easily be incorporated in the relative entropy term if we do not require the second argument to be a normalized probability distribution but instead allow arbitrary positive distributions. Our results extend to this more general setup but we will restrict to normalized distributions as otherwise the corresponding hypothesis testing problems are unnatural.

\subsubsection{Fixed marginal distribution}

Consider the following hypothesis testing problem:
\hypo{$\mathbb{H}^{\rm c}_{\ua}$}{$(X^n, Y^n) \sim P_{XY}^{\times n}$}{$X^n \sim R_X^{\times n}$ and $Y^n \sim P_Y^{\times n}$ are independent}
We find that the threshold rate is $D(P_{XY} \| R_X \times P_Y) = \log |\cX| - H(X|Y)$ and
the Hoeffding~\cite{hoeffding65} and Han-Kobyashi~\cite{han89} bounds establish an operational meaning for
 the R\'enyi conditional entropy
\begin{align}
  H_s^{\da}(X|Y) := \log |\cX| - D_s(P_{XY} \| R_X \times P_Y) 
  = \frac{1}{1 - s} \log \left( \sum_{y \in \cY} P_Y(y) \sum_{x \in \cX} P_{X|Y=y}(x)^{s} \right) \,.
\end{align}
This definition of conditional R\'enyi entropy is for example used to express the leaked modified mutual information in the secure random number generation~\cite[Thm.~2]{hayashi11b} and \cite[Thm.~2]{hayashitan15}.

\subsubsection{Arbitrary marginal distribution}
\label{sec:examples-ce-one}

We consider the following problem, in analogy with Section~\ref{sec:examples-mi-one}:
\hypo{$\mathbb{H}^{\rm c}_{\da}$}{$(X^n, Y^n) \sim P_{XY}^{\times n}$}{$X^n \sim R_X^{\times n}$ independent of $Y^n$\label{eq:hcond}}
This is covered by the example in Section~\ref{sec:ex1} with $T_X$ the uniform distribution. 
Again we determine the threshold $\log |\cX| - H(X|Y)$ and the error and strong converse exponents given operational significance to Arimoto's~\cite{arimoto75} definition of R\'enyi conditional entropy,
\begin{align}
  H_s^{\ua}(X|Y) := \log |\cX| - \min_{Q_Y \in \cQ(\cY)} D_{s}(P_{XY} \| R_X \times Q_Y) 
  = \frac{s}{1-s} \log \left( \sum_{y \in \cY} P_Y(y) \left( \sum_{x \in \cX} P_{X|Y=y}^{s} \right)^{\frac{1}{s}} \right) .
\end{align}
The minimum was evaluated using Sibson's identity (cf.\ Lemma~\ref{lm:sibson} in Appendix~\ref{sec:proof-examples}).

This definition has recently be reviewed in~\cite{fehr14} and compares favorably to other definitions of 
R\'enyi conditional entropy that have recently been put forward~\cite{teixeira12,iwamoto13}. For example, it has an operational interpretation determining the moments of the number of rounds required to guess $X$ from $Y$~\cite{arikan96}, and relatedly as an exponent for task encoding with side information~\cite{bunte14}.
More precsiely, we have 
\begin{align}
  \mathrm{e}_R(\mathbb{H}^{\rm c}_{\uda}) = \sup_{s \in (0,1)} \frac{1-s}{s} \big(\log |\cX|- H_s^{\uda}(X\!:\!Y) - R \big) 
  \quad \textrm{and} \quad
  \mathrm{sc}_R(\mathbb{H}^{\rm c}_{\uda}) 
= \sup_{s > 1} \frac{s-1}{s} \big( R -\log |\cX|+ H_s^{\uda}(X\!:\!Y) \big) ,
\end{align}
As a further example, in secure random number extraction,
$ \mathrm{sc}_{\log |\cX|-R}(\mathbb{H}^{\rm c}_{\uda})$
expresses the error exponent under the universal composability criterion~\cite[Thm.~4]{hayashi13} and \cite[Thm.~30]{hayashiwatanabe15}.

\subsection{R\'enyi Conditional Mutual Information: Testing Against Markov Distributions}
\label{sec:examples-cmi}

Let $X, Y$ and $Z$ be three random variables with a joint distribution $P_{XYZ} \in \cP(\cX \times \cY \times \cZ)$. The conditional mutual information, $I(X\!:\!Z|Y)$, can be seen as a measure of how close the distribution $P_{XYZ}$ is to a Markov chain $X \leftrightarrow Y \leftrightarrow Z$. For example, we can write
\begin{align}
  I(X\!:\!Z|Y) &= D(P_{XYZ} \| P_Y \times P_{X|Y} \times P_{Z|Y}) \label{eq:cmi1} \\
  &= \min_{Q_{Z|Y} \in \cP(Z|Y) } D(P_{XYZ} \| P_Y \times P_{X|Y} \times Q_{Z|Y}) \label{eq:cmi2} \\
   &= \min_{\substack{Q_{XYZ}  \in \cP(\cX \times \cY \times \cZ) \\ Q_{XYZ} = Q_{X|Y} \times Q_{Z|Y} \times Q_Z}} D(P_{XYZ} \| Q_Y \times Q_{X|Y} \times Q_{Z|Y}) \label{eq:cmi3} \,,
\end{align}
where the latter optimization is over all distributions satisfying the Markov condition $Q_{XYZ} = Q_{X|Y} \times Q_{Z|Y} \times Q_Z$.
These are only a few of all the possible expressions for the conditional mutual information, but we will focus our attention on these examples and follow a similar discussion as with the mutual information.

\subsubsection{All marginals fixed}
\label{sec:examples-cmi-1}

Again we first consider a simple alternative hypothesis. 
\hypo{$\mathbb{H}^{\mathrm{cmi}}_{\uuua}$}{$(X^n, Y^n, Z^n) \sim P_{XYZ}^{\times n}$}{$X^n \leftrightarrow Y^n \leftrightarrow Z^n$ is Markov, $(X^n, Y^n) \sim P_{XY}^{\times n}$, and $(Y^n, Z^n) \sim P_{YZ}^{\times n}$}

This corresponds to the sets $\cQ_n := \big\{ (P_{Y} \times P_{X|Y} \times P_{Z|Y} )^{\times n} \big\}$. The threshold rate for this problem is the conditional mutual information, $I(X\!:\!Z|Y)$.
Furthermore, the Hoeffding~\cite{hoeffding65} and Han-Kobyashi~\cite{han89} bounds yield an operational interpretation of the R\'enyi conditional mutual information given as
\begin{align}
  I_s^{\uuua}(X\!:\!Z|Y) &:= D_s(P_{XYZ} \| P_Y \times P_{X|Y} \times P_{Z|Y})  \\
  &\,= \frac{1}{s-1} \log \left( \sum_{y \in \cY} P_Y(y) \left(  \sum_{z \in \cZ} P_{Z|Y=y}(z) \left( \sum_{x \in \cX} P_{X|Y=y,Z=z}(x)^s P_{X|Y=y}(x)^{1-s} \right) \right) \right) \,.
\end{align}

In the special case where $P_{XYZ} = P_{YZ} \times P_{X|Z}$, this kind of R\'enyi mutual information describes the exponent for leaked mutual information when we employ a superposition code in the wire-tap channel~\cite[Lem.~16]{watanabe15} and~\cite[Thm.~20]{hayashi12b}.


\subsubsection{Two marginals fixed, recovery channels}
\label{sec:examples-cmi-rec} 
The expression in~\eqref{eq:cmi2} corresponds to the following problem:
\hypo{$\mathbb{H}^{\mathrm{cmi}}_{\uuda}$}{$(X^n, Y^n, Z^n) \sim P_{XYZ}^{\times n}$}{$X^n \leftrightarrow Y^n \leftrightarrow Z^n$ is Markov, $(X^n, Y^n) \sim P_{XY}^{\times n}$\label{eq:hcmi}}

This problem is discussed in Section~\ref{sec:ex3}. We show that the threshold for this test is again given by $I(X\!:\!Z|Y)$.
Moreover, the optimal error and strong converse exponents for $R$ close to the threshold are given by
\begin{align}
  \mathrm{e}_R(\mathbb{H}^{\rm cmi}_{\uuda}) = \sup_{s \in (0,1)} \frac{1-s}{s} \big( I_s^{\uuda}(X\!:\!Z|Y) - R \big) 
  \quad \textrm{and} \quad
  \mathrm{sc}_R(\mathbb{H}^{\rm cmi}_{\uuda}) = \sup_{s > 1} \frac{s-1}{s} \big( R - I_s^{\uuda}(X\!:\!Z|Y) \big) \,,
\end{align}
where we have introduced a new definition of the R\'enyi conditional mutual information. This is given by
\begin{align}
 I_s^{\uuda}(X\!:\!Z|Y) &:= \min_{Q_{Z|Y} \in \cP(\cZ|\cY)} D_s(P_{XYZ} \| P_Y \times P_{X|Y} \times Q_{Z|Y}) \\
 &\,= \frac{1}{s-1} \log \left( \sum_{y \in \cY} P_Y(y) \left( \sum_{z \in \cZ} P_{Z|Y=y}(z) \left( \sum_{x \in \cX} P_{X|Y=y,Z=z}(x)^s P_{X|Y=y}(x)^{1-s} \right)^{\!\frac{1}{s}} \right)^{\!\!s\,} \right) . \label{eq:cmi-closed-form}
\end{align}
The minimum was evaluated using Sibson's identity in Lemma~\ref{lm:sibson} for the distribution $Q_{Z|Y=y}$ seperately for each $y \in \cY$. (See Appendix~\ref{sec:proof-examples-3} for details.)
The resulting expression can be regarded as a conditional version of the Gallager function by replacing $s$ with $\frac{\rho}{\rho-1}$.
In the special case where $P_{XYZ} = P_{YZ} \times P_{X|Z}$, this quantity is used in superposition coding to describe the error exponent~\cite[Sec.~II]{kaspi11} and the exponent of leaked information~\cite[Thm.~22]{hayashi12b}.

\subsubsection{Arbitrary Markov distribution, permutation invariant}
\label{sec:examples-cmi-full}

The most general alternative hypothesis that we consider is comprised of all distributions that have a Markov structure $X \leftrightarrow Y \leftrightarrow Z$. More precisely, the following problem:
\hypo{$\mathbb{H}^{\mathrm{cmi}}_{\ddda}$}{$(X^n, Y^n, Z^n) \sim P_{XYZ}^{\times n}$}{$X^n \leftrightarrow Y^n \leftrightarrow Z^n$ is Markov, $(X^n,Y^n,Z^n)$ permutation invariant}
This is covered in Section~\ref{sec:ex3}.
The threshold is again $I(X\!:\!Z|Y)$
and the optimal error and strong converse exponents for $R$ close to the threshold are given by
\begin{align}
  \mathrm{e}_R(\mathbb{H}^{\rm cmi}_{\ddda}) = \sup_{s \in (\frac23,1)} \frac{1-s}{s} \big( I_s^{\ddda}(X\!:\!Z|Y) - R \big) 
  \quad \textrm{and} \quad
  \mathrm{sc}_R(\mathbb{H}^{\rm cmi}_{\ddda}) = \sup_{s > 1} \frac{s-1}{s} \big( R - I_s^{\ddda}(X\!:\!Z|Y) \big) \,,
\end{align}
with yet another R\'enyi conditional mutual information,
\begin{align}
   I_s^{\ddda}(X\!:\!Z|Y) &:= \min_{Q_{XYZ} \in \cP(\cX\times\cY\times\cZ) } D_s(P_{XYZ} \| Q_Y \times Q_{X|Y} \times Q_{Z|Y}) \,.
\end{align}
Note that we only have an operational interpretation of this quantity for $s > \frac23$, and, moreover, we do not know of a closed form expression.


\section{Proofs: Hoeffding Bound}
\label{sec:hoeffding}

The proof of Theorem~\ref{th:hoeffding} is split into two parts, achievability and optimality, which both rely on different Axioms. 


\subsection[Some Properties of phi(s)]{Some Properties of $\phi(s)$}
\label{sec:convex1}

Before we state our result, let us introduce some helpful notation. Recall that $\phi(s) := (s-1) D_{s}(P\|\cQ) = \log g_s(P\|\cQ)$ and define
\begin{align}
    \bar{\phi}_{s_0}(s) := (s-1) D_{s}(P\|\hat{Q}^{s_0}) = \log g_s(P\|\hat{Q}^{s_0}) \,,
\end{align}
where $\hat{Q}^{s_0} := Q_{\hat{\theta}^{s_0}}$.
Clearly, $\phi(s_0) = \bar{\phi}_{s_0}(s_0)$ by definition of $\hat{\theta}^s$ in Axiom~\ref{ax:para}. An important consequence of this Axiom is the following lemma, which shows that the first derivative of $\phi$ and $\bar{\phi}$ agree as well at $s = s_0$.
\begin{lemma}
 \label{lm:derivative}
  Assume Axiom~\ref{ax:para} holds on $(a,b)$. Then, for $s_0 \in (a,b)$, we have
  \begin{align}
    \frac{\d}{\d s} D_{s}(P\|\cQ) \Big|_{s = s_0} = \frac{\d}{\d s} D_{s}(P\|Q_{\hat{\theta}^{s_0}}) \Big|_{s=s_0} \,.
  \end{align}
  Moreover, the function $\phi(s)$ on $(a,b)$ is continuously differentiable and satisfies $\phi'(s_0) = \bar{\phi}_{s_0}'(s_0)$. 
\end{lemma}

\begin{IEEEproof}
  Let us define $f(s,\theta) = g_{s}(P\|Q_{\theta})$. Write $\theta = (\theta_1, \ldots, \theta_d)$ as a $d$-dimensional real vector. The point $\hat{\theta}^s$ is determined by the implicit functions
  \begin{align}
    F_i(s,\theta) := \frac{\partial}{\partial \theta_i} f(s,\theta) = 0 , \qquad \forall i \in \{1, 2, \ldots, d\} \,.
  \end{align}
  Moreover, the Hessian matrix $H_s$ of $\theta \mapsto f(s,\theta)$ at $\hat{\theta}^s$ is given by
  \begin{align}
    (H_s)_{i,j} = \frac{\partial^2}{\partial \theta_i \partial \theta_j} f(s,\theta) \big|_{\theta = \hat{\theta}^{s}} \,.
  \end{align}
  The map $s \mapsto H_s$ is continuous since $(s,\theta) \mapsto f(s,\theta)$ is $C^2$ by Axiom~\ref{ax:para}.
  
  Let us first treat the case $s_0 > 1$. By Axiom~\ref{ax:para} we have that $H_s$ is positive definite and thus invertible. 
  Then, the implicit function theorem yields
  \begin{align}
    \frac{ \partial \hat{\theta}_i^{s} }{\partial s} \bigg|_{s=s_0} = - \sum_{j=1}^d \big( H_{s_0}^{-1} \big)_{i,j} \cdot\frac{ \partial F_j(s,\hat{\theta}^{s_0}) }{\partial s} \bigg|_{s = s_0 } \label{eq:thetas}
  \end{align}
  and in particular $s \mapsto \hat{\theta}^s$ is continuously differentiable. 
  We further find that
  \begin{align}
    \frac{\d}{\d s} f(s, \hat{\theta}^s) \Big|_{s = s_0} 
    = \frac{\partial}{\partial s} f(s, \hat{\theta}^{s_0}) \Big|_{s = s_0} + \sum_{i=1}^d F^i(s_0,\hat{\theta}^{s_0}) \cdot \frac{ \partial \hat{\theta}_i^{s} }{\partial s} \bigg|_{s=s_0}
    = \frac{\partial}{\partial s} f(s, \hat{\theta}^{s_0}) \Big|_{s = s_0} \,,
  \end{align}
  where we used that $F^i(s_0,\hat{\theta}^{s_0}) = 0$ by definition of $\hat{\theta}^{s_0}$. This establishes the result for $s_0 \in (1,b)$. An analogous argument yields the same result for $s_0 < 1$. For $s_0 = 1$ we instead choose $f(s,\theta) = D_s(P\|Q_{\theta})$. Again the Hessian matrix for the derivative with regards to $\theta$ is strictly positive definite, and the remainder of the argument proceeds as before.
  
Finally, since $s \mapsto D_s(P\|Q)$ is $C^2$ for fixed $P, Q \in \cP(X)$ and $\theta \mapsto Q_{\theta}$ as well as $s \mapsto \hat{\theta}^s$ are continuous, we deduce that the functions $s \mapsto D_s(P\|\cQ)$ and $\phi(s) = (s-1) D_{s}(P\|\cQ)$ are continuously differentiable. 
\end{IEEEproof}

\subsection[Some Properties of Convex Continuously Differentiable Functions]{Some Properties of Convex $C^1$ Functions}
\label{sec:convex2}
\newcommand{\phit}{\tilde{\phi}}
\newcommand{\psit}{\tilde{\psi}}

In this section let $\phit(s)$ be a general convex $C^1$ function on $(a, b) \subseteq [0, \infty)$ and define $\psit(s) := s \phit'(s) - \phit(s)$. Moreover, define $\tilde{R}_{c} := \lim_{s \to c} \psit(s)$ for every $c \in [a, b]$.

\begin{lemma}
\label{lm:hmono}
  The function $\psit(s)$ is continuous and monotonically nondecreasing on $(a,b)$.
\end{lemma}
\begin{IEEEproof}
  Let $a < s_0 < s_1 < b$. 
  By the mean value theorem there exists an $s \in [s_0, s_1]$ such that $(s_1 - s_0) \phit'(s) = \phit(s_1) - \phit(s_0)$. Thus, 
  \begin{align}
    \psit(s_1) - \psit(s_0) &= s_1 \phit'(s_1) - s_0 \phit'(s_0)  - \big ( \phit(s_1) - \phit(s_0) \big) \\
    &= s_1 \phit'(s_1) - s_0 \phit'(s_0) - (s_1 - s_0) \phit'(s) \\
    &= s_1 \big( \phit'(s_1) - \phit'(s) \big) + s_0 \big( \phit'(s) - \phit'(s_0) \big) \geq 0 \,.
  \end{align}
  The inequality follows from the assumption that $\phit(s)$ is convex, and $\phit'(s)$ thus monotonically increasing.
\end{IEEEproof}

\begin{lemma}
\label{lm:supmax}
  Let $R \in (\tilde{R}_a, \tilde{R}_b)$. Then, there exists an $\hat{s} \in (a, b)$ such that $\psit(\hat{s}) = R$ and
  \begin{align}
    \sup_{s \in (a,b)} \frac{(s-1)R - \phit(s)}{s} = \frac{(\hat{s} - 1)R - \phit(\hat{s})}{\hat{s}}
  \end{align}
\end{lemma}

\begin{IEEEproof}
 By continuity, for every $R \in (\tilde{R}_a, \tilde{R}_b)$, there exists (at least one) value $\hat{s} \in (a, b)$ such that $\psit(\hat{s}) = R$. 
  Let us first calculate the derivative of $g(s) := \frac{(s-1) R - \phit(s)}{s}$. This yields
  \begin{align}
   g'(s)\,   = 
  \frac{s R - s \phit'(s) - (s-1) R + \phit(s) }{s^2}
  = \frac{ R - \psit(s) }{s^2} \,.
  \end{align}
  Note that the numerator is monotonically decreasing in $s$ due to Lemma~\ref{lm:hmono} and vanishes at $s = \hat{s}$. In particular, we find that $g'(s) \geq 0$ for $a < s < \hat{s}$ and $g'(s) \leq 0$ for $\hat{s} < s < b$. We conclude that $\hat{s}$ maximizes $g(s)$ on $(a, b)$.
\end{IEEEproof}


\subsection{Proof of Achievability}

Achievability for Theorem~\ref{th:hoeffding} follows from the following statement.
\begin{prop}
  \label{pr:hoeffding-achieve}
  Assume Axioms~\ref{ax:add} and~\ref{ax:universal} hold with parameter $a$. Then, we have
  \begin{equation}
  \label{eq:hoeffding-achieve}
    \liminf_{n \to \infty} - \frac{1}{n} \log 
\hat{\alpha}\Big(\exp(-n R);P^{\times n} \Big\| \,\overline{\cQ}_n \Big) 
  \geq \sup_{s \in (a, 1)} \left\{ \frac{1-s}{s} \big( D_{s}(P\|\cQ) - R \big) \right\}. 
  \end{equation}
\end{prop}

\begin{IEEEproof}
Note that the expression on the right hand side of~\eqref{eq:hoeffding-achieve} is zero if $R \geq  D(P\|\cQ)$ and the inequality thus holds trivially for that case. We assume that $R < D(P\|\cQ)$ for the remainder of this proof.

Let us fix any $s \in (a,1)$ for the moment. 
Moreover, let us define the sequence of tests
  \begin{align}
T_n(x^n) := 
\begin{cases} 
1 & \textrm{ if } P^{\times n}(x^n) \geq \exp(\lambda_n) U^n(x^n) \\
0 & \textrm{ otherwise}
\end{cases}
 , \label{eq:thetest}
\end{align}
where $U^n$ is the universal distribution of Axiom~\ref{ax:universal}.
We also choose the sequence $\{\lambda_{n} \}_{n \in \mathbb{N}}$ of real numbers 
as
\begin{align}
  \lambda_n = \frac{1}{s} \Big( \log v(n) + n R + (s-1) D_{s} (P^{\times n} \| U^n) \Big) . 
  \label{eq:sc-mu22}
\end{align}

Axiom~\ref{ax:universal} implies that $\cQ_n$ is closed under symmetrization. Moreover, the test $T_n$ is permutation invariant. Hence, for all $\pi \in S_n$, we have $Q^n[T_n] = Q^n[T_n W_{\pi}^n] = Q^n W_{\pi^{-1}}^n[T^n]$ and we can in particular replace $Q^n$ with its symmetrization. This yields
\begin{align}
  \beta(T_n; \cQ_n)
&= \max_{Q^n \in \cQ_n} Q^n \left[ P^{\times n}(X^n) 
\geq \exp(\lambda_n) U^n(X^n) \right] \label{eq:optimize-linear} \\
&= \max_{Q^n \in \cQ_n^{\sym}} Q^n \left[ P^{\times n}(X^n) 
\geq \exp(\lambda_n) U^n(X^n) \right] \,. \label{eq:optimize-sym}
\end{align}
Next we use the universal distribution in Axiom~\ref{ax:universal} to further bound
\begin{align}
\beta(T_n; \cQ_n)
&\leq v(n) \sum_{x^n \in \cX^n} \!\! U^n(x^n) \, 1 \left\{ P^{\times n}(x^n) \geq \exp(\lambda_n) U^n(x^{n}) \right\} \\
&\leq v(n) \exp(-s \lambda_n) \sum_{x^n \in \cX^n} \!\! \big(U^n(x) \big)^{1-s} \big( P^{\times n}(x) \big)^{s} \, 1 \left\{ P^{\times n}(x^n) \geq \exp(\lambda_n) U^n(x^{n}) \right\} \\
&\leq v(n) \exp(-s \lambda_n) \sum_{x^n \in \cX^n} \!\! \big(U^n(x) \big)^{1-s} \big( P^{\times n}(x) \big)^{s} \\
&= v(n) \exp\left(-s \lambda_n\right) \exp\left( (s-1) D_{s}( P^{\times n} \| U^n ) \right) . 
\end{align}
Hence, the requirement that 
$\beta(T_n; \cQ_n)  \leq \exp(-n R)$ is satisfied by the choice of $\lambda_n$ in~\eqref{eq:sc-mu22}. Note that this statement can directly be extended to the convex hull due to~\eqref{eq:convex-hull}.

Let us now take a closer look at the error of the first kind. 
Using a similar development as above, we find
\begin{align}
\hat{\alpha}\big(\exp(-n R); P^{\times n} \big\|\overline{\cQ}_n \big) 
\leq \alpha(T_n; P^{\times n}) 
&= P^{\times n} \left[ P^{\times n}(X^n) < \exp(\lambda_n) U^n(X^{n}) \right] \\
&\leq \exp\big( (1-s) \lambda_n  \big) \exp \big( (s-1) D_{s} (P^{\times n}\| U^n) \big)  \\
&= \exp \Big( \frac{1-s}{s} \big( \log v(n) + nR - D_{s} (P^{\times n}\| U^n) \big) \Big)
\end{align}
where we substituted $\lambda_n$ from~\eqref{eq:sc-mu22} in the last step. Further using the additivity property of Axiom~\ref{ax:add}, we find that
\begin{align}
  D_{s} (P^{\times n}\| U^n) \geq D_{s} (P^{\times n} \| \cQ_n) \geq n D_s(P \| \cQ),
\end{align}
and thus we arrive at the bound
\begin{align}
 \log \hat{\alpha}\big(\exp(-n R); P^{\times n} \big\|\overline{\cQ}_n \big) \leq \frac{1-s}{s} \big( nR - nD_s(P \| \cQ) + \log v(n) \big)  \label{6-22-1} \,.
  \end{align}
  Since $\log v(n) = O(\log n)$, taking the limit $n \to \infty$ yields
    \begin{align}
     \liminf_{n \to \infty}  - \frac{1}{n} \log 
\hat{\alpha}\big(\exp(-n R); P^{\times n} \big\|\overline{\cQ}_n\big) 
 \geq 
\frac{1-s}{s} \big(D_{s}(P\|\cQ) - R\big) \,.
  \end{align}
Finally, since this derivation holds for all $s \in (a,1)$, we established the direct part.
\end{IEEEproof}

\subsection{Proof of Optimality}

To show optimality, we will directly employ the converse of the Hoeffding bound.

\begin{prop}
  Assume Axioms~\ref{ax:prod}--\ref{ax:para} hold with parameter $a$. Then, for any $R > R_a$, we have
  \begin{equation}
  \label{eq:hoeffding-conv}
    \limsup_{n \to \infty} - \frac{1}{n} \log 
\hat{\alpha}\Big(\exp(-n R);P^{\times n} \Big\| \cQ_n \Big) 
  \leq \sup_{s \in (a, 1)} \left\{ \frac{1-s}{s} \big( D_{s}(P\|\cQ) - R \big) \right\}. 
  \end{equation}
\end{prop}

\begin{IEEEproof}
    Let us first consider the case $R \in (R_a, D(P\|\cQ))$.  
We will use the results of Sections~\ref{sec:convex1} and~\ref{sec:convex2}. Take $\hat{s} \in (a, 1)$ to be the optimizer in Lemma~\ref{lm:supmax} for the functions $\phit = \phi$ and $\psit = \psi$ on $(a,1)$. Then we have that $\psi(\hat{s}) = \hat{s} \phi'(\hat{s}) - \phi(\hat{s}) = R$ and
\begin{align}
  \sup_{s \in (a, 1)} \frac{1-s}{s} \left( D_{s}(P \| \cQ) - R \right) = \frac{1-\hat{s}}{\hat{s}} \left( D_{\hat{s}}(P \| \cQ) - R \right) \label{eq:supmax1}
\end{align}
The following consequence of Lemma~\ref{lm:derivative} is crucial. Recall that
\begin{align}
  \bar{\phi}(s) = \log g_s(P\|Q_{\hat{\theta}^{\hat{s}}}) \qquad \textrm{and} \qquad \bar{\psi}(s) = s \bar{\phi}'(s) - \bar{\phi}(s) \,, \label{eq:phibardef}
\end{align}
where $\hat{\theta}^{\hat{s}}$ is the optimal $\theta$ at $s = \hat{s}$. Note that the function $s \mapsto \bar{\phi}(s)$ is a $C^1$ convex function on all of $(0,1)$ since $g_s$ is evaluated for fixed distributions $P$ and $Q_{\hat{\theta}^{\hat{s}}}$.
Moreover, Lemma~\ref{lm:derivative} implies that $\phi'(\hat{s}) = \bar{\phi}'(\hat{s})$, and hence, 
$\bar{\psi}(\hat{s}) = \psi(\hat{s}) = R$. 
We can thus also apply Lemma~\ref{lm:supmax} for the functions $\phit = \bar{\phi}$ and $\psit = \bar{\psi}$ on $(0,1)$ and find that 
\begin{align}
  \sup_{s \in (0, 1)} \frac{1-s}{s} \left( D_{s}(P \| Q_{\hat{\theta}^{\hat{s}}}) - R \right) = \frac{1-\hat{s}}{\hat{s}} \left( D_{\hat{s}}(P \| \cQ) - R \right) \,. \label{eq:supmax2}
\end{align}

Next we note that, due to Axiom~\ref{ax:prod}, we have
\begin{align}
  \hat{\alpha}\Big(\exp(-n R);P^{\times n} \Big\| \cQ_n \Big) 
\geq   \hat{\alpha}\Big(\exp(-n R); P^{\times n} \Big\| Q^{\times n}\Big) 
\end{align}
  for any $Q \in \cQ$. 
  After applying this for $Q = Q_{\hat{\theta}^{\hat{s}}}$, we can further apply the converse of the Hoeffding bound (we take the formulation in~\cite[Thm.~1]{nagaoka06}) to the expression on the right-hand side, which yields
  \begin{align}
    &\limsup_{n \to \infty} \left\{ - \frac{1}{n} \log \hat{\alpha}\Big(\exp(-n R); P^{\times n} \Big\| \cQ_n \Big) \right\} \label{eq:lhs1} \\
    &\qquad \qquad \leq 
	\limsup_{n \to \infty} \left\{ - \frac{1}{n} \log \hat{\alpha}\Big(\exp(-n R); P^{\times n} \Big\| Q_{\hat{\theta}^{\hat{s}}}^{\times n}\Big) \right\}\\
    &\qquad \qquad = 
    \sup_{s \in (0, 1)} \frac{1-s}{s} \left( D_{s}(P \| Q_{\hat{\theta}^{\hat{s}}}) - R \right) .
\end{align}
  which proves the result together with~\eqref{eq:supmax1} and~\eqref{eq:supmax2}.
  
  If $R \geq D(P\|\cQ)$ the right hand side of~\eqref{eq:hoeffding-conv} evaluates to zero. Moreover, there exists at least one $Q \in \cQ$ such that $R \geq D(P\|Q)$. The result then follows from the converse of Hoeffding's bound since
  \begin{align}
    \limsup_{n \to \infty} - \frac{1}{n} \log 
\hat{\alpha}\Big(\exp(-n R);P^{\times n} \Big\| \cQ_n \Big) 
  \leq \limsup_{n \to \infty} - \frac{1}{n} \log 
\hat{\alpha}\Big(\exp(-n R);P^{\times n} \Big\| Q^{\times n} \Big) = 0 \,.
  \end{align}
\end{IEEEproof}


\section{Proofs: Strong Converse Exponents}
\label{sec:sc}

Again we treat achievability and optimality with separate proofs that rely on different axioms.

\subsection{Proof of Achievability}

Our proof relies on a variant of the G\"artner-Ellis theorem of large deviation theory (see, e.g.,~\cite[Sec.~2 and Sec.~3.4]{dembo98} for an overview), which we recall here. Given a sequence of random variables $\{ Z_n \}_{n \in \mathbb{N}}$ we
introduce its asymptotic \emph{cumulant generating function} as
\begin{align}
   \Lambda_Z(t) &:= \lim_{n \to \infty} \left\{ \frac{1}{n} \log \big( \mathbb{E} \left[ \exp ( n t Z_n ) \right] \big) \right\} ,
   \label{eq:cum}
\end{align}
if it exists.
For our purposes it is sufficient to use the following variant of the G\"artner-Ellis theorem due to Chen~\cite[Thm.~3.6]{chen00} (see also~\cite[Lem.~A.2]{mosonyi14} for this exact statement).
\begin{lemma}
  \label{pr:mosonyi}
  Let us assume that $t \mapsto \Lambda_Z(t)$ as defined in~\eqref{eq:cum} exists and is differentiable in some interval $(a, b)$. Then, for any $z \in \big( \lim_{t \searrow a} \Lambda_Z'(t), \lim_{t \nearrow b} \Lambda_Z'(t) \big)$, we have
  \begin{align}
    \limsup_{n \to \infty} \left\{ - \frac{1}{n} \log \Pr [ Z_n \geq z] \right\} \leq \sup_{t \in (a, b)} \left\{ t z - \Lambda_Z(t) \right\} \,.
  \end{align}
\end{lemma}

Achievability follows from the following statement.

\begin{prop}
  Assume Axioms~\ref{ax:add}--\ref{ax:para} hold with parameter $b$. For any $R \in (0, R_b)$, we have
  \begin{equation}
  \label{eq:sc-achieve}
    \lim_{n \to \infty} - \frac{1}{n} \log \bigg( 1 -
\hat{\alpha}\Big(\exp(-n R);P^{\times n} \Big\|\, \overline{\cQ}_n \Big) \bigg) 
  \leq \sup_{s \in (1, b)} \left\{ \frac{s-1}{s} \big( R - D_{s}(P\|\cQ) \big) \right\}. 
  \end{equation}
\end{prop}

\begin{IEEEproof}
  Let us first assume that $R \in (D(P\|\cQ), R_b)$. Using Lemma~\ref{lm:supmax} with $\phit = \phi$ on $(1,b)$, we find the value $\hat{s} \in (1,b)$ that satisfies
  \begin{align}
    \sup_{s \in (1,b)} \frac{s-1}{s} \big( R - D_{s}(P\|\cQ) \big) = \frac{\hat{s}-1}{\hat{s}} \big( R - D_{\hat{s}}(P\|\cQ) \big) \label{eq:defshat}
  \end{align}
  and $\psi(\hat{s}) = \hat{s} \phi'(\hat{s}) - \phi(\hat{s}) = R$. We use the same sequence of tests $T_n$ as in~\eqref{eq:thetest} and the sequence $\lambda_n$ of~\eqref{eq:sc-mu22}, substituting $\hat{s}$ for $s$. This ensures that $\beta(T_n; \overline{\cQ}_n) \leq \exp(-nR)$, as shown in the proof of Proposition~\ref{pr:hoeffding-achieve}. Moreover,
  \begin{align}
    1 - \hat{\alpha}\Big(\exp(-n R);P^{\times n} \Big\|\, \overline{\cQ}_n \Big) 
    &\geq 1 - \alpha(T_n; P^{\times n}) \label{eq:inserthere} \\
    &= P^{\times n} \left[ P^{\times n}(X^n) \geq \exp(\lambda_n) U^n(X^n) \right] = \Pr[Z_n \geq 0] \,,
  \end{align}
  where we defined the sequence of random variables $Z_n(X^n)$ following the law $X^n \leftarrow P^{\times n}$ and
  \begin{align}
    Z_n(x^n) 
    &= \frac{1}{n} \left( \log \frac{P^{\times n}(x^n)}{U^n(x^n)} - \lambda_n  \right) \\
    &= \frac{1}{n} \left( \log \frac{P^{\times n}(x^n)}{U^n(x^n)} 
       - \frac{\log v(n)}{\hat{s}} - \frac{\hat{s}-1}{\hat{s}} D_{\hat{s}}(P^{\times n}\|U^n) - \frac{nR}{\hat{s}} \right) \,. \label{eq:Zn}
  \end{align}
  Its asymptotic cumulant generating function then evaluates to
  \begin{align}
       \Lambda_Z(t-1) =&\ \lim_{n \to \infty} \left\{ \frac{1}{n} \log \big( \mathbb{E} \left[ \exp ( n (t-1) Z_n ) \right] \big) \right\} \\
       =&\ \lim_{n \to \infty} \left\{ \frac{1}{n} \log \mathbb{E} \left[ \frac{P^{\times n}(X^n)^{t-1}}{U^n(X^n)^{t-1}} \right] - (t-1) \left( \frac{\log v(n)}{n \hat{s}} + \frac{\hat{s}-1}{n\hat{s}} D_{\hat{s}}(P^{\times n}\|U^n) + \frac{R}{\hat{s}} \right) \right\} \\
       =&\ (t-1) \lim_{n \to \infty} \left\{ \frac{1}{n} D_{t}(P^{\times n}\| U^n) - \frac{\hat{s}-1}{n\hat{s}}  D_{\hat{s}}(P^{\times n}\|U^n) - \frac{R}{\hat{s}} - \frac{\log v(n)}{n \hat{s}} \right\} \\
       =&\ \phi(t) - \frac{t-1}{\hat{s}} ( \phi(\hat{s}) + R ) \label{eq:arriveat} \\
       =&\ \phi(t) - (t-1) \phi'(\hat{s}) \,.
  \end{align}
Here we used the expression for $Z_n$ given in~\eqref{eq:Zn} and the defintion of the R\'enyi divergence in the second and third equality, respectively. To arrive at~\eqref{eq:arriveat} we evaluated the limit $n \to \infty$ using Lemma~\ref{lm:universal-limit}. The last equality follows from the relation between $\hat{s}$ and $R$ given in the line following~\eqref{eq:defshat}.

Note that $t \mapsto \Lambda_Z(t-1)$ is differentiable on $(1, b)$ due to Lemma~\ref{lm:derivative}. Moreover, in order to apply Lemma~\ref{pr:mosonyi} with $z=0$, we need to verify the following two inequalities:
  \begin{align}
    \lim_{t \to 1} \left\{ \frac{\d}{\d t} \Lambda_Z(t-1) \right\}
    = \phi'(1) - \phi'(\hat{s}) < 0 \quad \textrm{and} \qquad
    \lim_{t \to b} \left\{ \frac{\d}{\d t} \Lambda_Z(t-1) \right\}
    = \phi'(b) - \phi'(\hat{s}) > 0 \,. \label{eq:ineq}
  \end{align}
  We cannot invoke strict convexity of $\phi(s)$ to verify the above bounds; instead, note that $D(P\|\cQ) < R$, and thus
  \begin{align}
    \phi'(1) - \phi'(\hat{s}) &= D(P\|\cQ) - \frac{\hat{s}-1}{\hat{s}} D_{\hat{s}}(P\|\cQ) - \frac{1}{\hat{s}} R \\
    &< \frac{\hat{s}-1}{\hat{s}} \big( D(P\|\cQ) - D_{\hat{s}}(P\|\cQ) \big) \leq 0 \,.
  \end{align}
  To prove the second inequality in~\eqref{eq:ineq}, we use the fact that $R_b > R$ to show 
  \begin{align}
    \phi'(b) - \phi'(\hat{s}) &= \frac{1}{b} R_b + \frac{b-1}{b} D_b(P\|\cQ) - \frac{1}{\hat{s}} R - \frac{\hat{s}-1}{\hat{s}} D_{\hat{s}}(P\|\cQ) \\
    &> \frac{1}{b} R + \frac{b-1}{b} D_b(P\|\cQ) - \frac{1}{\hat{s}} R - \frac{\hat{s}-1}{\hat{s}} D_{\hat{s}}(P\|\cQ) \\
    &= \frac{\hat{s}-1}{\hat{s}} \big( R - D_{\hat{s}}(P\|\cQ) \big) - \frac{b-1}{b} \big( R - D_b(P\|\cQ) \big) \geq 0 \,,
  \end{align}
  where the last inequality follows by the definition of $\hat{s}$.
  
  We have now verified the conditions of Lemma~\ref{pr:mosonyi} with $z=0$, which yields
  \begin{align}
    \lim_{n \to \infty} -\frac{1}{n} \log \Pr[Z_n \geq 0] &= \sup_{t \in (1,b)} (t-1)\phi'(\hat{s}) - \phi(t) \label{eq:supneeded}\\
    &= (\hat{s}-1)\phi'(\hat{s}) - \phi(\hat{s})  \label{eq:supdone}\\
    &= \frac{\hat{s}-1}{\hat{s}} \bigg( \hat{s} \phi'(\hat{s}) - \phi(\hat{s}) + \frac{1}{\hat{s}-1} \phi(\hat{s}) \bigg) = \frac{\hat{s}-1}{\hat{s}} \big( R - D_{\hat{s}}(P\|\cQ) \big) \,.
  \end{align}
  To evaluate the supremum in~\eqref{eq:supneeded}, we note that the objective function $t \mapsto (t-1)\phi'(\hat{s}) - \phi(t)$ is concave in $t$ and its derivative vanishes at $t = \hat{s}$. This establishes~\eqref{eq:supdone}. Combining this with~\eqref{eq:inserthere} concludes the proof.

  For $R \leq D(P\|\cQ)$ the right hand side of~\eqref{eq:sc-achieve} evaluates to zero. Since the expression on the left hand side is clearly monotonically increasing in $R$ we deduce that, for all such $R$, 
  \begin{align}
    \lim_{n \to \infty} - \frac{1}{n} \log \bigg( 1 -
\hat{\alpha}\Big(\exp(-n R);P^{\times n} \Big\|\, \overline{\cQ}_n \Big) \bigg) 
  \leq \inf_{R > D(P\|\cQ)} \sup_{s \in (1, b)} \left\{ \frac{s-1}{s} \big( R - D_{s}(P\|\cQ) \big) \right\} = 0 \,.
  \end{align}

\end{IEEEproof}

\subsection{Proof of Optimality}

Optimality follows as a corollary of Han and Kobayashi's~\cite{han89} derivation of the strong converse exponent.
\begin{prop}
Assume Axiom~\ref{ax:prod} holds.
For any $R \geq 0$, we have
  \begin{equation}
    \liminf_{n \to \infty} - \frac{1}{n} \log \bigg( 1 -
\hat{\alpha}\Big(\exp(-n R);P^{\times n} \Big\| \cQ_n \Big) \bigg)
  \geq \sup_{s > 1} \left\{ \frac{s-1}{s} \big( R - D_{s}(P\|\cQ) \big) \right\}.
  \end{equation}
\end{prop}

\begin{IEEEproof}
If $R < D(P\|\cQ)$ the bound holds trivially. Otherwise,
analogous to the optimality proof for Theorem~\ref{th:hoeffding}, we first fix $Q \in \cQ$ and apply the Han-Kobayashi converse bound~\cite{han89} (in the form of~\cite[Ch.~VI]{ogawa00} and~\cite[Thm.~IV.9]{mosonyiogawa13}). This yields
  \begin{align}
&\liminf_{n \to \infty} \left\{ - \frac{1}{n} \log \bigg( 1 -
	\hat{\alpha}\Big(\exp(-n R);P^{\times n} \Big\| \cQ_n \Big) \bigg) \right\} \\
&\qquad \quad \geq \liminf_{n \to \infty} \left\{ - \frac{1}{n} \log \bigg( 1- \hat{\alpha}\Big(\exp(-n R); 	P^{\times n} \Big\| Q^{\times n} \Big) \bigg) \right\}\\
    &\qquad \quad = 
\sup_{s > 1} \left\{ \frac{s-1}{s} \big( R - D_{s}(P \Big\| Q) \big) \right\} \label{eq:rhs2}.
  \end{align}
  As this holds for all $Q \in \cQ$, we maximize the expression in~\eqref{eq:rhs2} over $Q$ to arrive at the desired result.
\end{IEEEproof}


\section{Proofs: Second Order Asymptotics of Stein's Lemma}
\label{sec:second}

In the proofs of this section we assume that $\log$ denotes the natural logarithm for ease of presentation.
The following result refines Lemma~\ref{lm:universal-limit}, and is the key ingredient of our proof.

\begin{lemma} \label{lm:universal2}
Assume Axioms~\ref{ax:prod}--\ref{ax:para} holds on $(a,b) \supset \{ 1 \}$. For all $t \in \mathbb{R}$, 
  \begin{align}
    \lim_{n \to \infty} \left\{ \frac{t}{\sqrt{n}} \Big( D_{1 + \frac{t}{\sqrt{n}}}(P^{\times n} \| U^n) - n D(P\|\cQ) \Big) \right\} = \frac{t^2}{2} V(P\|\cQ) \,. \label{eq:uni2}
  \end{align}
\end{lemma}

\begin{IEEEproof}
  By combining Lemma~\ref{lm:derivative} at $s_0 = 1$ with the Taylor expansion in~\eqref{eq:taylor}, we find $D_{1+s}(P\|\cQ) = D(P\|\cQ) + \frac{s}{2} V(P\|\cQ) + O(s^2)$.
  From~\eqref{eq:step1} we learn that
  \begin{align}
    D_{1+\frac{t}{\sqrt{n}}}(P^{\times n}\| U^n) 
    \leq n D_{1+\frac{t}{\sqrt{n}}}(P\|\cQ) + \log v(n)
    = n D(P\|\cQ) + \frac{t\sqrt{n}}{2} V(P\|\cQ) + O(\log n) \,.  \label{eq:second1}
  \end{align}
  Furthermore, employing additivity from Axiom~\ref{ax:add} yields
  \begin{align}
  D_{1+\frac{t}{\sqrt{n}}}(P^{\times n}\| U^n) 
    \geq n D_{1+\frac{t}{\sqrt{n}}}(P\|\cQ)
    = n D(P\|\cQ) + \frac{t\sqrt{n}}{2} V(P\|\cQ) + O(\log n) \,. \label{eq:second2}
   \end{align}
   Combining~\eqref{eq:second1} and~\eqref{eq:second2} yields the desired statement.
\end{IEEEproof}

Now let $M_{X}(t) := \mathbb{E}\big[ \exp(t X) \big]$ denote the moment generating function of a real random variable $X$. We also need the following property of moment generating functions, a variant of L\'evi's continuity theorem~\cite[Thm.~2]{mukherjea06}.
\begin{lemma}\label{lm:moments}
   Let $0 < a < b$. If a sequence of random variables $\{ X_n \}_{n \in \mathbb{N}}$ satisfies $\lim_{n \to \infty} M_{X_n}(t) = M_X(t)$ for some random variable $X$ and all $t \in (a, b)$, then $\lim_{n \to \infty} \Pr[X_n \leq k] = \Pr[X \leq k]$ for all $k \in \mathbb{R}$.
\end{lemma}

We prove the direct and converse part of Theorem~\ref{th:second} together.

\begin{IEEEproof}[Proof of Theorem~\ref{th:second}]
  We first show the converse statement. Choosing the optimal distribution $\hat{Q} = \hat{Q}^1 \in \cQ$ as defined in~\eqref{eq:mini2},
  we find
  \begin{align}
    \hat{\alpha}\big( \exp(- n D(P \|\cQ ) - \sqrt{n}r) ; P^{\times n} \big\| \cQ_n \big) \geq \hat{\alpha}\big( \exp(- n D(P \|\hat{Q} ) - \sqrt{n}r) ; P^{\times n} \big\| \hat{Q}^{\times n} \big) ,
  \end{align}
  and the limiting statement then follows using~\cite[Thm.\ 2]{yushkevich53} (see also~\cite[Thm.\ 1.1]{strassen62}).
  
  To show achievability we again rely on the test given in~\eqref{eq:thetest} and set $\lambda_n = n D(P \| \cQ) + \sqrt{n} r + \log v(n)$. Then by Axiom~\ref{ax:universal} and using the argument leading to~\eqref{eq:optimize-sym} to establish the first identity, we have
  \begin{align}
    \beta(T_n; \cQ_n) &= \max_{Q^n \in \cQ_n^{\sym}} Q^n\big[ P^{\times n}(X^n) \geq \exp(\lambda_n) U^n(X^n) \big] \\
    &\leq v(n) U^n \big[ P^{\times n}(X^n) \geq \exp(\lambda_n) U^n(X^n) \big] \\
    &\leq v(n) \exp(-\lambda_n) P^{\times n} \big[ P^{\times n}(X^n) \geq \exp(\lambda_n) U^n(X^n)\big] \\
    &\leq \exp \big(- n D(P\|\cQ) - \sqrt{n} r \big) \,.
  \end{align}
  
  Furthermore, we find
  \begin{align}
    \alpha(T_n; P^{\times n}) &= P^{\times n} \big[ \log P^{\times n}(X^n) - \log U^n(X^n) < n D(P\|\cQ) + \sqrt{n} r + \log v(n) \big] \\
    &= \Pr [ Y_n(X^n) < r] \,,
  \end{align}
  where $X^n \sim P^{\times n}$ and we defined the following sequence of random variables as
  \begin{align}
    Y_n(X^n) := \frac{1}{\sqrt{n}} \big( \log P^{\times n}(X^n) - \log U^n(X^n) - n D(P\|\cQ) - \log v(n) \big) \,.
  \end{align}
  
  Lemma~\ref{lm:universal2} then implies that the cumulant generating function converges to
  \begin{align}
    \log M_Y(t) &= \lim_{n \to \infty} \log \mathbb{E}\big[ \exp(t Y^n) \big] \\
    &= \lim_{n \to \infty} \left\{ \log \mathbb{E}\big[ P_{X^n}(x^n)^{\frac{t}{\sqrt{n}}} U^n(x^n)^{-\frac{t}{\sqrt{n}}} \big]  + \sqrt{n} t D(P \| \cQ) - \frac{t}{\sqrt{n}} \log v(n) \right\} \\
     &= \lim_{n \to \infty} \left\{ \frac{t}{\sqrt{n}} \big( D_{1+\frac{t}{\sqrt{n}}}(P^{\times n}\|U^n) - n D(P \|\cQ) - \log v(n) \big) \right\} \\
    &= \frac{t^2}{2} V(P\|\cQ) \,.
  \end{align}
  Hence, by Lemma~\ref{lm:moments}, the sequence of random variable $\{ Y_n \}_n$ converges in distribution to a random variable $Y$ with cumulant generating function $\log M_Y(t)$, i.e., a Gaussian random variable with zero mean and variance~$V(P\|\cQ)$. In particular, this yields
\begin{align}
\lim_{n \to \infty} P^{\times n} \left[ Y_n < r \right] 
= \Pr \left[ Y < r \right] 
= \Phi \left( \frac{r}{\sqrt{V(P\|\cQ)}} \right) . 
\label{eq:sec2}
\end{align}
Since $\hat{\alpha}\big( \exp(- n D(P \|\cQ ) - \sqrt{n}r) ; P^{\times n} \big\| \cQ_n \big) \leq \alpha(T_n; P^{\times n})$, this concludes the proof.
\end{IEEEproof}



\section{Conclusion}
\label{sec:conc}

We have introduced a general framework to treat binary hypothesis testing with a composite alternative hypothesis. In this general framework we show analogues of Stein's Lemma, Hoeffding's optimal error exponents and Han-Kobyashi's optimal strong converse exponents. 
We have discussed several concrete examples that lead to operational
interpretations of various R\'enyi information measures.

The coincidence between our obtained exponents for the hypothesis testing problem in~\eqref{eq:hchannel} and the corresponding exponents for channel coding is quite interesting. A similar coincidence has been observed for the case of source coding with side information and~\eqref{eq:hcond}. These facts seem to indicate a deep relation between 
coding and the composite alternative hypotheses given in \eqref{eq:hchannel} and \eqref{eq:hcond}.
Its further clarification is an interesting future direction of study, for example one could try to find coding problems that are closely related to~\eqref{eq:hcmi}.

In statistics, the $\chi^2$ test is used in an asymptotic setting similar to \eqref{eq:hcomp}.
The test assumes i.i.d.\ distributions and is used for the case when both hypotheses are composite (see, e.g.,~\cite{lehmann05}).
For small samples and $k=2$, Fisher's exact test~\cite{fisher22} can be used to replace the $\chi^2$ test.
Recently, the setting of small samples and general $k$ has been studied using a Gr\"obner basis approach~\cite{diaconis98,sakata05}.
In contrast to their formulation, we have not assumed i.i.d.\ structure for the independent case; instead, we only require permutation invariance for each random variable.
Our result suggests that we can replace the i.i.d.\ condition by a permutation invariant condition
when testing independence, which can be expected to have wider applications.

The key ingredient of our derivation is an axiomatic approach based on the universal distribution. Due to its generality, we can treat many composite hypothesis testing problems
without i.i.d.\ assumption (for the composite hypothesis), and it will be interesting to explore
further examples that fit into our framework.
Moreover, because the universal distribution plays an important role in universal channel coding~\cite{hayashi09}, we can expect that it will play an important role when analyzing universal protocols for other problems in information theory.

As explained in Section~\ref{sec:ex2}, we cannot remove the permutation invariance condition in that example, an essential difference to the example given in Section~\ref{sec:ex1}.
This kind of difference sheds light on the difference between channel coding and secure random number generation.
Originally, for the channel coding, the meta converse was introduced using simple hypothesis testing~\cite[Sec.~3]{nagaoka01} and~\cite{polyanskiy10}. Polyanskiy~\cite[Sec.~II]{polyanskiy13} then extended it to the composite hypothesis testing of the form~\eqref{eq:hpoly}.
Although this improvement does not effect the exponents and the second-order coding rate, it can improve the bound in the finite blocklength regime.
Recently, Tyagi-Watanabe~\cite{tyagi14,tyagi14b} introduced a converse bound for secure random number generation by using simple hypothesis testing between a true joint distribution and an arbitrary product distribution.
Although in their converse bound, we can choose an arbitrary product distribution as the alternative hypothesis, we cannot replace the alternative hypothesis by a composite hypothesis composed of all of product distributions. Hence, for secure random number generation, we cannot extend their bound to a bound based on the composite hypothesis as in~\cite{polyanskiy13}.

In prior work~\cite{hayashitomamichel14} the present authors have analyzed composite hypothesis testing in the non-commutative (quantum) regime and found an operational interpretation for various definitions of quantum R\'enyi mutual information and quantum R\'enyi conditional entropy~\cite{tomamichel13}. However, the present work is more general than the classical specialization of that work and requires new techniques. This allows us to deal with more complex composite alternative hypotheses, and in particular allows for a characterization of R\'enyi conditional mutual information.
Furthermore, finding appropriate definitions for R\'enyi conditional mutual information in the non-commutative setting is an ongoing topic of research~\cite{bertawilde14}. It is possible that an adaption of our analysis to quantum hypothesis testing will lead to further progress in this direction. However, some caution is advised since already the  definitions of the regular conditional mutual information in~\eqref{eq:cmi1}--\eqref{eq:cmi3} are not equivalent in the quantum case.

\paragraph*{Acknowledgements}
We thank Vincent Y.~F.~Tan for helpful comments and Christoph Pfister for alterting us to several typos and small gaps in our presentation.
MH is partially supported by a MEXT Grant-in-Aid for Scientific Research (A) No. 23246071, and by the National Institute of Information and Communication Technology (NICT), Japan.
MT is funded by an University of Sydney Postdoctoral Fellowship and acknowledges support from the ARC Centre of Excellence for Engineered Quantum Systems (EQUS). 



\bibliographystyle{ultimate}
\bibliography{library}

\begin{IEEEbiographynophoto}{Marco Tomamichel} (M'13--SM'16) is a Senior Lecturer at the Centre for Quantum Software and Information and School of Software, Faculty of Engineering and Information Technology at the University of Technology Sydney. He received a M.Sc.\ in Electrical Engineering and Information Technology degree from ETH Zurich (Switzerland) in 2007. He then graduated with a Ph.D.\ in Physics at the Institute of Theoretical Physics at ETH Zurich in 2012. Before commencing his current position, he has been a Senior Research Fellow at the Centre for Quantum Technologies at the National University of Singapore and a Lecturer at the School of Physics at the University of Sydney.
  
  His research interests include classical and quantum information theory with finite resources as well as applications to cryptography.
  \end{IEEEbiographynophoto}

\begin{IEEEbiographynophoto}{Masahito Hayashi}(M'06--SM'13--F'17) was born in Japan in 1971.
  He received the B.S. degree from the Faculty of Sciences in Kyoto
  University, Japan, in 1994 and the M.S. and Ph.D. degrees in Mathematics from
  Kyoto University, Japan, in 1996 and 1999, respectively. He worked in Kyoto University as a Research Fellow of the Japan Society of the
  Promotion of Science (JSPS) from 1998 to 2000,
  and worked in the Laboratory for Mathematical Neuroscience,
  Brain Science Institute, RIKEN from 2000 to 2003,
  and worked in ERATO Quantum Computation and Information Project,
  Japan Science and Technology Agency (JST) as the Research Head from 2000 to 2006.
  He also worked in the Superrobust Computation Project Information Science and Technology Strategic Core (21st Century COE by MEXT) Graduate School of Information Science and Technology, The University of Tokyo as Adjunct Associate Professor from 2004 to 2007.
  He worked in the Graduate School of Information Sciences, Tohoku University as Associate Professor from 2007 to 2012.
  In 2012, he joined the Graduate School of Mathematics, Nagoya University as Professor.
  He also worked in Centre for Quantum Technologies, National University of Singapore as Visiting Research Associate Professor from 2009 to 2012
  and as Visiting Research Professor from 2012 to now.
  In 2011, he received Information Theory Society Paper Award (2011) for ``Information-Spectrum Approach to Second-Order Coding Rate in Channel Coding''.
  In 2016, he received the Japan Academy Medal from the Japan Academy
  and the JSPS Prize from Japan Society for the Promotion of Science.
  
  In 2006, he published the book ``Quantum Information: An Introduction''  from Springer, whose revised version was published as ``Quantum Information Theory: Mathematical Foundation'' from Graduate Texts in Physics, Springer in 2016.
  In 2016, he published other two books ``Group Representation for Quantum Theory'' and ``A Group Theoretic Approach to Quantum Information'' from Springer.
  He is on the Editorial Board of {\it International Journal of Quantum Information}
  and {\it International Journal On Advances in Security}.
  His research interests include classical and quantum information theory and classical and quantum statistical inference.
  \end{IEEEbiographynophoto}

\appendices

\section{Verification of the Axioms for Examples in Section~\ref{sec:examples}}
\label{sec:proof-examples}

\subsection{Sibson's identity}
\label{sec:sibson}

This appendix proves that the examples satisfy our axioms. One of the main ingredients is Sibson's identity~\cite{sibson69}, as presented in~\cite[Eq.~(11)--(13)]{csiszar95}.
\begin{lemma}
\label{lm:sibson}
For any distributions $P_{XY} \in \cP(\cX \times \cY)$, $T_X \in \cP(\cX)$ and $Q_Y \in \cP(\cY)$, and any $s \in (0,1) \cup (1, \infty)$,
\begin{align}
  D_{s}(P_{XY} \| T_X \times Q_Y) = D_{s}(P_{XY} \| T_X \times \hat{Q}_Y^s) + D_{s}(\hat{Q}_Y^s \| Q_Y) \, \label{eq:sibson}
\end{align}
where the optimal distribution $\hat{Q}_Y^s \in \cP(\cY)$ is given by
\begin{align}
 \hat{Q}_Y^s(y) = \frac{P_Y(y) g_s(P_{X|Y=y}\| T_X)^{\frac1s}}{\sum_y P_Y(y) g_s(P_{X|Y=y}\| T_X)^{\frac1s}} \,.
\end{align}
Thus, in particular, $\argmin_{Q_Y \in \cP(\cY)} D_{\alpha}(P_{XY} \| T_X \times Q_Y) = \{ Q_Y^* \}$.
\end{lemma}

\begin{IEEEproof}
  We rewrite~\eqref{eq:sibson} as $g_{s}(P_{XY} \| T_X \times Q_Y) = g_s(P_{XY} \| T_X \times \hat{Q}_Y^s) \cdot g_{s}(\hat{Q}_Y^s \| Q_Y)$, at which point the equality can be verified by close inspection. The fact that $\hat{Q}_Y^s$ is the unique minimizer is then a consequence of the positive definiteness of the R\'enyi divergence.
\end{IEEEproof}

\subsection{Proof of Proposition~\ref{pr:examples-one}}
\label{sec:proof-examples-1}

\begin{IEEEproof}
  Clearly $\cQ$ is compact convex and we explicitly find the optimizer using Sibson's identity in Lemma~\ref{lm:sibson}. Up to normalization it is given by
  \begin{align}
    \hat{Q}_Y^s(y) &\sim P_Y(y)  g_s ( P_{X|Y=y} \| T_X )^{\frac1s},  \label{eq:Qopt}
  \end{align}
  and thus Axiom~\ref{ax:convex} is verified. Axiom~\ref{ax:prod} holds by definition and Axiom~\ref{ax:add} can be verified by noting that $\hat{Q}^s$ in~\eqref{eq:Qopt} takes on an i.i.d.\ product form when both $P_{XY}^{\times n}$ and $T_X^{\times n}$ are i.i.d.\ products.
  
  Next, note that $\cQ_n$ is closed under permutations and convex. The universal distributions are
  \begin{align}
    U_{X^nY^n}^n = T_X^{\times n} \times U_{Y^n}^n, \qquad \textrm{with} \qquad U_{Y^n}^n(y^n) = \sum_{\lambda \in \cT_n(\cY)} \frac{1}{| \cT_n(\cY) |} \frac{ 1 }{ |\lambda| } \, 1 \{ y^n \textrm{ is of type } \lambda \} \, ,
  \end{align}
  as in Lemma~\ref{lm:uni-dist}. Clearly $U_{X^nY^n}^n \in \cQ_n^{\sym}$ and thus.
  we find that Axiom~\ref{ax:universal} is satisfied with $v(n) = |\cT_n(Y)| = \poly(n)$.
  
  Finally note that all the above remains true if we restrict $\cQ_n$ to permutation invariant or i.i.d.\ product distributions, denoted $\cQ_n'$, except that now $U_{X^nY^n}^n \notin \cQ_n'$. However, we still have
  \begin{align}
     D_s(P_{XY}^{\times n} \| U_{X^nY^n}^n) \geq D_s(P_{XY}^{\times n} \| \cQ_n) = D_s(P_{XY}^{\times n} \| \cQ_n') ,
  \end{align}
  since additivity property guarantees that the minimum in $D_s(P_{XY}^{\times n} \| \cQ_n)$ is taken by a product distribution.
\end{IEEEproof}

\subsection{Proof of Proposition~\ref{pr:examples-two}}
\label{sec:proof-examples-2}

  We give the proof for the case $k=1$ and set $\cX_1 = \cX$. The generalization to larger $k$ does not require further conceptual insights, and we will remark in a footnote where nontrivial changes are necessary.

\begin{IEEEproof}
  Axiom~\ref{ax:prod} holds by definition. To verify Axiom~\ref{ax:universal} we first note that
  the joint permutations of $X^n$ and~$Y^n$ separate as
  $W_{X^nY^n}^n[\pi] = W_{X^n}^n[\pi] \times W_{Y^n}^n[\pi]$. Thus, we can write
  \begin{align}
    \sum_{\pi \in S_n} \frac{1}{n!} \big(Q_{X^n} \times Q_{Y^n} \big) W_{X^nY^n}^n[\pi] = \sum_{\pi \in S_n} \frac{1}{n!} Q_{X^n} W_{X^n}^n[\pi] \times Q_{Y^n} W_{Y^n}^n[\pi] = Q_{X^n} \times \sum_{\pi \in S_n} \frac{1}{n!}  Q_{Y^n} W_{Y^n}^n[\pi] \,,
  \end{align}
  where we used that $Q_{X^n} \in \cP^{\sym}(\cX^n)$ to establish the last equality. Clearly the resulting distribution lies in $\cQ_n$. Next, consider the universal distribution $U_{X^nY^n}^n = U_{X^n}^n \times U_{Y^n}^n$ with $U_{X^n}^n$ and $U_{Y^n}^n$ given as in~\eqref{eq:universal1}. Clearly, since $\cQ_n^{\sym} = \cP^{\sym}(\cX^n) \times \cP^{\sym}(\cY^n)$ we then find that every symmetric distribution satisfies
\begin{align}
  Q_{X^n} \times Q_{Y^n} \leq |\cT_n(\cX)|  U_{X^n}^n \times |\cT_n(\cY)| U_{Y^n}^n
\end{align}
and Axiom~\ref{ax:universal} holds with $v(n) = |\cT_n(\cX)| |\cT_n(\cY)| = \poly(n)$.

  For Axiom~\ref{ax:para} we chose the following parametrization. Since $\cP(\cX)$ is convex subset of $\mathbb{R}^{|\cX|-1}$, there exists a natural smooth parametrization $\Theta_1 \ni \theta_1 \mapsto Q_{X,\theta_1} \in \cP(\cX)$ where $\Theta_1$ is a convex subset of $\mathbb{R}^{|\cX|-1}$, and similarly for $\cP(\cY)$. Combining these two parameterizations, we introduce a $\Theta \subset \mathbb{R}^{|\cX| + |\cY| - 2}$ such that 
  \begin{align}
     \Theta \ni \theta = (\theta_1, \theta_2) \mapsto Q_{X,\theta_1} \times Q_{Y,\theta_2} = Q_{XY,\theta} \in \cQ \,.
  \end{align}
  The set $\Theta$ is evidently convex. 
  Let us next verify the required convexity and concavity properties. First note that the map $f(x,y) = x^{1-s} y^{1-s}$ for $x, y \geq 0$ is strictly jointly concave when $s \in (\frac12, 1)$ and strictly jointly convex when $s > 1$. This follows, for example, from studying the Hessian matrix for $f$, which is
\begin{align}
  \left(\begin{matrix} \frac{\partial^2 f}{\partial x^2} & \frac{\partial^2 f}{\partial x\partial y} \\ \frac{\partial^2 f}{\partial y\partial x} & \frac{\partial^2 f}{\partial y^2} \end{matrix} \right) 
  = x^{-s}y^{-s} (1-s) \left(\begin{matrix} -s x^{-1} y & (1-s) \\ (1-s) & -s x y^{-1} \end{matrix} \right) 
\end{align}
  and its determinant $x^{-2s}y^{-2s} (1-s)^2 ( 2s - 1)$. The Hessian is negative definite for $s \in (\frac12, 1)$ and positive definite for $s > 1$ when $x,y > 0$.\footnote{For $k \geq 2$ we need to consider the function $f(x_1, x_2, \ldots, x_k, y) = \prod_{i=1}^k x_i^{1-s} y^{1-s}$ and the range of $s$ where this function is jointly concave in all its arguments is further restricted to $s > \frac{k}{k+1}$.} 
  From this and the above parametrization we conclude that the Hessian of the map $\Theta \mapsto g_s(P_{XY}\|Q_X \times Q_Y)$ is negative definite for $s \in (\frac12, 1)$ and positive definite for $s > 1$ in the relative interior of $\Theta$.
  Moreover, the function $g(x,y) = -\log xy = -\log x - \log y$ evidently has a positive definite Hessian for $x, y > 0$.
  Hence the desired concavity and convexity properties for the maps $\theta \mapsto g_s(P_{XY}\|Q_{XY,\theta})$ and $\theta \mapsto D_s(P_{XY}\|Q_{XY,\theta})$ hold. As a consequence, the minimizer is unique if it exists in the relative interior of $\cQ$ (as we will show below).
    
Let us assume that 
  $D_s(P_{XY} \| \cQ) = D_s(P_{XY} \| \hat{Q}_X^s \times \hat{Q}_Y^s)$ for some optimal distributions $\hat{Q}_X^s$ and $\hat{Q}_Y^s$. To verify Axiom~\ref{ax:add} we note that distributions are optimal only if they satisfy the self-consistency relation for a local optimum  given in Eq.~\eqref{eq:Qopt}. This implies that
  \begin{align}
      \hat{Q}_X^s(x) \sim P_X(x) g_s ( P_{Y|X=x} \| \hat{Q}_Y^s )^{\frac{1}{s}} \quad \textrm{and} \quad 
    \hat{Q}_Y^s(y) \sim P_Y(y) g_s ( P_{X|Y=y} \| \hat{Q}_X^s )^{\frac{1}{s}} \,. \label{eq:self-consistency}
  \end{align}
  This solution is in the relative interior of $\cQ$ and thus unique (for $\alpha > \frac12$).
  
  Furthermore, we find that the product distributions $\hat{Q}_X^s \times \hat{Q}_X^s$ and $\hat{Q}_Y^s \times \hat{Q}_Y^s$ satisfy the self-consistency relations for the local optimal solution of $D_s(P_{XY}^{\times 2} \| \cQ_2)$. More precisely, we find
  \begin{align}
    \hat{Q}_X^s(x_1)\hat{Q}_X^s(x_2) \sim P_X(x_1)P_X(x_2) g_s ( P_{Y|X=x_1} \times P_{Y|X=x_2} \| \hat{Q}_Y^s \times \hat{Q}_Y^s)^{\frac{1}{s}} ,
  \end{align}
  and vice versa. Moreover, due to the fact that the function is either convex or concave for $\alpha > \frac12$, we can conclude that these product distributions are globally optimal as well. Namely
  \begin{align}
   D_{s}(P_{XY}^{\times 2} \| \cQ_2 ) = D_{s}\big(P_{XY}^{\times 2} \big\| (\hat{Q}_{X} \times \hat{Q}_{Y})^{\times 2}\big) = 2 D_s(P_{XY} \| \cQ) \,.
  \end{align}
  Applying this argument inductively yields the condition of Axiom~\ref{ax:add}.
\end{IEEEproof}

\subsection{Proof of Proposition~\ref{pr:examples-three}}
\label{sec:proof-examples-3}

Before we commence with the proof we need to introduce some additional concepts and auxiliary results.
Let us introduce the following representation of channels, which is reminiscent of the Choi-Jamio{\l}kowski isomorphism in quantum information theory.  
Let $\cX$ and $\cY$ be finite sets. For any channel $Q_{Y|X} \in \cP(\cY|\cX)$ we define a vector representation $\tilde{Q}_{XY} = 1_{X} \times Q_{Y|X}$, where $1_X$ is the identity vector for the Schur (element-wise) product of vectors, i.e.\ $1_{X}(x) = 1$ for all $x \in \cX$. More concretely, the vector is given by $\tilde{Q}_{XY}(x,y) = Q_{Y|X}(y|x)$. Note that $\tilde{Q}_{XY}$ is not a probability distribution but clearly we must have
\begin{align}
  \label{eq:normalize-choi}
  \sum_{y \in \cY} \tilde{Q}_{XY}(x,y) = 1, \qquad \forall x \in \cX \,. 
\end{align}
Using this representation we can write joint distribution after the application of the channel as
\begin{align}
  P_{X} \times Q_{Y|X} = (P_{X} \times 1_{Y}) \circ \tilde{Q}_{XY} ,  \label{eq:schur}
\end{align}
where $\circ$ denotes the Schur product between the vectors and $1_{Y}(y) = 1$ for all $y \in \cY$. Note also that the normalization condition~\eqref{eq:normalize-choi} enforces that the resulting vector is a probability distribution, and hence every vector with positive elements that satisfies~\eqref{eq:normalize-choi} corresponds to a valid channel.

\begin{lemma} 
\label{lm:choi-covariance}
  $Q_{Y^n|X^n}$ is covariant under permutations if and only if $\tilde{Q}_{X^nY^n}$ is permutation invariant. Formally,
  \begin{align}
       \forall \pi \in S_n:\ Q_{Y^n|X^n} W_{Y^n}[\pi] = W_{X^n}[\pi] Q_{Y^n|X^n}  
       \quad \iff \quad \forall \pi \in S_n:\ \tilde{Q}_{X^nY^n} W_{X^nY^n}[\pi] = \tilde{Q}_{X^nY^n} \,.
  \end{align}
\end{lemma}

\begin{IEEEproof}
  The following equalities can be verified by close inspection:
  \begin{align}
   \tilde{Q}_{X^nY^n} W_{X^nY^n}[\pi] &= \big(1_{X^n} \times Q_{Y^n|X^n} \big) \big( W_{X^n}[\pi] \times W_{Y^n}[\pi] \big) \\
   &= ( 1_{X^n} W_{X^n}[\pi] ) \times \big( W_{X^n}[\pi^{-1}] Q_{Y^n|X^n} W_{Y^n}[\pi] \big) \\
   &= 1_{X^n} \times \big( W_{X^n}[\pi^{-1}] Q_{Y^n|X^n} W_{Y^n}[\pi] \big) \,.
   \end{align}
   The equivalence of the two conditions then follows from the fact that $W_{X^n}[\pi^{-1}] W_{X^n}[\pi]$ is the identity channel.
\end{IEEEproof}

\begin{IEEEproof}[Proof of Lemma~\ref{lm:uni-channel}]
  Let $Q_{Y^n|X^n}$ be covariant under permutations. Then from Lemma~\ref{lm:choi-covariance} we learn that $\tilde{Q}_{X^nY^n}$ is permutation invariant and thus in particular can be written in the form
  \begin{align}
    \tilde{Q}_{X^nY^n}(x^n,y^n) = \!\! \sum_{\lambda_{XY} \in \cT_n(\cX\times\cY)} \frac{\tilde{q}_{XY}(\lambda_{XY})}{\# \lambda_{XY}} 1\{ (x^n,y^n) \textrm{ is of type } \lambda_{XY} \} \, , \label{eq:theform}
  \end{align}
  where $\tilde{q}_{XY}$ is a probability distribution over joint types $\lambda_{XY}$ and  $\#\lambda_{XY}$ denotes the number of sequences of type $\lambda_{XY}$. Moreover, Eq.~\eqref{eq:normalize-choi} enforces that for every type $\mu_X \in \cT_n(\cX)$, and any sequence $x^n$ of type $\mu_X$, we have
  \begin{align}
    1 = \sum_{y^n \in \cY^n} \tilde{Q}_{X^nY^n}(x^n,y^n)
    &= \sum_{\lambda_{XY} \in \cT_n(\cX\times\cY) } \!  \frac{\tilde{q}_{XY}(\lambda_{XY})}{\# \lambda_{XY}} \underbrace{ \sum_{y^n \in \cY^n}  1\{ (x^n,y^n) \textrm{ is of type } \lambda_{XY} \} }_{ =\  \# \lambda_{Y|X}(x^n) }\,, \label{eq:condition}
  \end{align}
  where the number of sequences of type $\lambda_{XY}$ with marginal $x^n$, denoted $\# \lambda_{Y|X}(x^n)$, clearly only depend on the type $\mu_X$ of the marginal. Moreover, if the type of $x^n$ does not correspond to the marginal type $\lambda_X$ of $\lambda_{XY}$ then $\# \lambda_{Y|X}(x^n)$ vanishes. Generally, we have
$\# \lambda_{Y|X}(x^n) = 1\{ \mu_x = \lambda_x \} \frac{\# \lambda_{XY} }{ \# \lambda_{X}}$.
 Hence~\eqref{eq:condition} simplifies to
  \begin{align}
    \sum_{\substack{\lambda_{XY} \in \cT_n(\cX\times\cY) \\ \lambda_X = \mu_X }}  \tilde{q}_{XY}(\lambda_{XY}) \frac{1}{\# \lambda_{X}} = 1
    \qquad \forall \mu_X \in \cT_n(\cX) \,. \label{eq:thecondition}
  \end{align}
  A direct consequence of this condition is that $\tilde{q}_{XY}(\lambda_{XY}) \leq \# \lambda_X$ for all $\lambda_{XY}$.
  
  Now let us define a universal permutation covariant channel $\tilde{U}_{X^nY^n}$ by the choice
  \begin{align}
    \tilde{u}_{XY}(\lambda_{XY}) := \frac{\# \lambda_{X}}{ \big| \{ \kappa_{XY} \in \cT_n(\cX\times\cY) : \kappa_X = \lambda_X \} \big|}
  \end{align}
  which evidently satisfies~\eqref{eq:thecondition}. Moreover, for all permutation covariant channels with representation $\tilde{Q}_{XY}$ of the form~\eqref{eq:theform}, we have the bound
  \begin{align}
    \tilde{q}_{XY}(\lambda_{XY}) \leq \# \lambda_X \leq \big| \{ \kappa_{XY} \in \cT_n(\cX\times\cY) : \kappa_X = \lambda_X \} \big| \tilde{u}_{XY}(\lambda_{XY}) \leq | \cT_n(\cX\times\cY) | \tilde{u}_{XY}(\lambda_{XY}) \,. 
  \end{align}
  Hence also $\tilde{Q}_{XY} \leq | \cT_n(\cX\times\cY) |\, \tilde{U}_{XY}$. The statement of the lemma then follows from the expression in~\eqref{eq:schur}.
\end{IEEEproof}

\begin{IEEEproof}[Proof of Proposition~\ref{pr:examples-three}]
  The set $\cQ$ is clearly compact convex. Note that
  \begin{align}
    g_s(P_{XYZ} \| P_Y \times P_{X|Y} \times Q_{Z|Y} ) = \sum_{y \in \cY} P_Y(y) \, g_s (P_{XZ|Y=y} \|  P_{X|Y=y} \times Q_{Z|Y=y}) \,.
  \end{align}
  From this we can then deduce, as in~\eqref{eq:Qopt}, that the optimal channel takes on the form
    \begin{align}
    \hat{Q}_{Z|Y=y}^s(z) 
    \sim P_{Z|Y=y}(z) g_s \big( P_{X|Y=y,Z=z} \big\| T_{X|Y=y} \big)^{\frac1s},  \label{eq:Qopt2}
  \end{align}
for all $y \in \cY$. As such, it is clear that Axioms~\ref{ax:convex}--\ref{ax:add} are satisfied.
  
It remains to verify Axiom~\ref{ax:universal}. First note that $\cQ_n$ is closed under symmetrization. Moreover, any channel $Q_{Z^n|Y^n}$ corresponding to a permutation invariant element of $\cQ_n^{\sym}$ satisfies
\begin{align}
   Q_{Z^n|Y^n} W_{Z^n}[\pi] = W_{Y^n}[\pi] Q_{Z^n|Y^n}\,, \qquad \forall \pi \in S_n ,
\end{align}
i.e.\ $Q_{Z^n|Y^n}$ is permutation covariant. Hence, Lemma~\ref{lm:uni-channel} applies and guarantees the existence of a sequence of universal channels $\{U^n_{Z^n|Y^n}\}_{n\in\mathbb{N}}$ with $U^n_{Z^n|Y^n} \in \cQ_n^{\sym}$ such that 
\begin{align}
   P_{XY}^{\times n} \times Q_{Z^n|Y^n}(x^n,y^n,z^n) \leq v(n) P_{XY}^{\times n} \times U^n_{Z^n|Y^n}(x^n,y^n,z^n) \qquad \forall x^n \in \cX^n, y^n \in \cY^n, z^n \in \cZ^n \,.
\end{align}
\end{IEEEproof}

\subsection{Proof of Proposition~\ref{pr:examples-four}}
\label{sec:proof-examples-4}

\begin{IEEEproof}
The proof proceeds similarly to the proofs of Propositions~\ref{pr:examples-two} and~\ref{pr:examples-three}. Axiom~\ref{ax:prod} holds by definition and Axiom~\ref{ax:universal} can be verified using the
universal distributions
\begin{align}
  U_{X^nY^nZ^n}^n = U_{Y^n}^n \times U_{X^n|Y^n}^n \times U_{Z^n|Y^n}^n
\end{align}
with the universal distributions $U_{Y^n}^n$ as in~\eqref{eq:universal1} and the universal maps $U_{X^n|Y^n}^n$ and $U_{Z^n|Y^n}^n$ provided by Lemma~\ref{lm:uni-channel}. 

Axiom~\ref{ax:para} is verified with a construction analogous to Proposition~\ref{pr:examples-two} but this time we need to consider the function $(x,y,z) \mapsto x^{1-s}y^{1-s}z^{1-s}$ which is strictly jointly concave for $s \in (\frac23, 1)$ and strictly jointly convex for $s > 1$.
Finally, Axiom~\ref{ax:add} is again verified using the self-consistency relations.
\end{IEEEproof}

\end{document}